\documentclass[aps,prb,superscriptaddress,nofootinbib,showpacs,twocolumn]{revtex4}
\usepackage{float}
\usepackage{amssymb}
\usepackage{graphicx}
\usepackage{epsf,epsfig}
\usepackage{bm}
\usepackage{bbm}
\usepackage{amsfonts}
\usepackage{amsmath}
\usepackage{graphics}
\usepackage[normalem]{ulem}
\usepackage{color}

\newcommand{\be}{\begin{eqnarray}}
\newcommand{\ee}{\end{eqnarray}}
\newcommand{\ba}{\begin{array}}
\newcommand{\ea}{\end{array}}

\newcommand{\bfr}{{\bf r}}

\begin{document}
\title{Flexural phonons in supported graphene: from pinning to localization}
\begin{abstract}
We identify graphene layer on a disordered substrate as a system where localization of phonons can be observed. Generally, observation of localization for scattering waves is not simple, because the Rayleigh scattering is inversely proportional to a high power of wavelength. The situation is radically different for the out of plane vibrations, so-called flexural phonons, scattered by pinning centers induced by a substrate. In this case, the scattering time for vanishing wave vector tends to a finite limit. One may, therefore, expect that the physics of the flexural phonons exhibits features characteristic for electron localization in two dimensions, albeit without complications caused by the electron-electron interactions. We confirm this idea by calculating statistical properties of the Anderson localization of flexural phonons for a model of elastic sheet in the presence of the pinning centers. Finally, we discuss possible manifestations of the flexural phonons, including the localized ones, in the electronic thermal conductance.
\end{abstract}

\author{Wei L.Z. Zhao}
\affiliation{Department of Physics \& Astronomy, Texas A\&M University, College Station, TX 77843-4242, USA}

\author{Konstantin S. Tikhonov}
\affiliation{L. D. Landau Institute for Theoretical Physics, 117940 Moscow, Russia}
\affiliation{Skolkovo Institute of Science and Technology, 143026 Skolkovo, Russia}

\author{Alexander M. Finkel'stein}
\affiliation{Department of Physics \& Astronomy, Texas A\&M University, College Station, TX 77843-4242, USA}
\affiliation{Department of Condensed Matter Physics, The Weizmann Institute of Science, 76100 Rehovot, Israel}

\date{\today}

\maketitle



\noindent Most of the research in graphene emphasizes the relativistic character of its electron spectrum. However, graphene is also interesting due to its out-of-plane (flexural) vibrational phonon modes.  Flexural phonons (FPs) are a unique addition that van der Waals heterostructures have brought into microscopic physics. \cite{GeimGrigorieva} 
Usually, FPs are considered in the context of the suspended graphene.  
Here we argue that graphene layer placed on the top of the \emph{supporting} 
SiO$_2$ substrate 
gives an opportunity to observe Anderson localization \cite{Abrahamsbook} 
for the FPs.

To get an idea, let us recall the known facts about scattering of a long-wave acoustic wave by a cylinder. The result 
depends drastically on the boundary conditions for the velocity potential $\Phi$ on the surface of the cylinder.  
\cite{morse19861953methods}
If the velocity component 
normal to the surface of the cylinder vanishes, \emph{i.e.,} $\left.\partial_r\Phi\right|_{r=a}=0$, the scattering cross-section $\sigma_R$ is proportional to $a(ka)^3 $, where $a$ is the radius of the cylinder and $k$ is the wave vector. This is the conventional Rayleigh scattering result \cite{rayleigh1896theory}
for two dimensional $(2d)$ geometry. However, when pressure is constant, the boundary condition reads $\Phi(a)=0$, and this 
influences 
the scattering substantially. 
Unlike the Rayleigh scattering, the zero angular harmonic is involved, and as a result the cross section diverges at small $k$ as $\propto (k\ln^{2} \frac{1}{ka})^{-1}$. (The same takes place for an electro-magnetic wave scattering by a metallic cylinder.)

In graphene, the substrate cannot scatter effectively the usual acoustic waves, longitudinal and transverse, because graphene itself is one of the most rigid substances. 
The situation with the out-of-plane vibrations is quite different. From the analysis of intrinsic and extrinsic corrugation of monolayer graphene deposited on SiO$_2$ substrate,
it has been concluded that in this system the layer is suspended between hills of the substrate landscape.
\cite{ishigami2007nanoletters,stolyarova2007PNAS,geringer2009intrinsic,leroy2009spatially}
We have checked that scattering of the FPs by areas attached 
to the substrate is similar to the scattering by a rigid obstacle. \cite{norris1995scattering}
The zero harmonic is also involved, and the scattering cross-section 
diverges as 
$\sigma_{fl}= 4/k$. 

In this work we study statistical properties of the out of plane excitations for a pinned-suspended flexible sheet. Whether pinning centers are located in the vicinity of the maximal heights of the substrate where the interaction with the layer 
is the strongest, or there are charges on the substrate which interact strongly with their images, will be not important for our purposes. First of all, we are interested in the scattering rate of the FPs in the presence of the randomly located pinning centers with concentration $n_i$.  
Taking into consideration that the spectrum of the FPs 
$\omega(k)=\alpha k^2$ is quadratic, \emph{i.e.,} velocity is linear in $k$, one obtains a scattering rate $\tau^{-1}=v\sigma_{fl}n_i$, that is finite in the low-energy limit.
Thus, 
for the FPs one may expect localization of the \emph{low-energy} modes with $\omega(k)<\tau^{-1}$. This is in striking contrast with localization of acoustic modes, which is known to happen only at high enough frequency. \cite{john1983localization, kirkpatrick1985localization, akkermans1985weak,sepehrinia2008numerical, monthus2010anderson} 

\bigskip
Let us comment upon the graphene layer deposited on the top of the corrugated substrate. Naively, the membrane-like layer either follows the substrate or hovers over the surface at some distance. However, measurements 
with the use of cantilevers \cite{conley2011graphene} 
point toward a possibility of the detaching a graphene
sheet from the substrate to relieve its strain by slipping.
(This is manifest by straightening of the cantilever.) In the case of the SiO$_2$ substrate, both experiment and theory agree that for typical magnitude of the corrugations, the graphene layer 
is partially detached from the substrate. 
Moreover, the theoretical considerations \cite{sabio2008electrostatic,kusminskiy2011pinning} 
justify the use of a contact force that is finite when graphene is conforming to the substrate and zero
otherwise.
The basic experimental facts \cite{geringer2009intrinsic} which lead to the conclusion 
that graphene layer deposited on SiO$_2$
is partly freely suspended are as follows: 
The long-range corrugation of the substrate with the correlation length of about 25\emph{nm} is also visible on the graphene sheet, but with a smaller amplitude than on the substrate. Mesoscopic corrugations with smaller length of about 15\emph{nm} not induced by the substrate were also identified. These short range corrugations are similar in height and wavelength to the ones observed in suspended graphene. \cite{meyer2007suspended,laitinen2014} 
In addition, the picture of partially suspended graphene, and the presence of the FPs in the graphene on the SiO$_2$ substrate, has been confirmed 
by the transport \cite{morozov2008giant} and thermal measurements. \cite{seol2010two,pumarol2012imaging}

\bigskip
\noindent \textbf{RESULTS}

\noindent Here, we demonstrate that the FPs in 
a pinned-suspended flexible sheet are more similar to disordered electrons rather than to acoustical phonons.
The main point here is 
that 
the pinning centers are effectively rigid obstacles for the FPs. As we have already explained, this leads to the non-vanishing scattering rate at small energies. 
Let us touch upon this point in more detail.

\bigskip
\noindent \textbf{Pinning potential as a barrier for FPs} 

\noindent Usually by a rigid obstacle one understands an inclusion with the Young's modulus much higher than that in surrounding area. For graphene, which itself is very rigid, this is not an issue. However, the pinning potential introduces an energy barrier 
of a finite height
for the flexural modes: 
\be
\label{deq}
\kappa {\nabla}^4 h(\bfr,t)+\rho \frac{\partial^2 h(\bfr,t)}{\partial t^2}=-\rho {\omega_0(\bfr)}^2 h(\bfr,t),
\ee
Here $h(\bfr,t)$ is displacement in the out of plane direction; the term describing the barrier is on the right-hand side of the above equation. As a result, a flexural phonon with an energy smaller than $\omega_0$ cannot enter the area of pinning. We have checked that such an inclusion is equivalent to a rigid obstacle. For not small $ka$, the cross-section $\sigma_{fl}(k)=4f(ka)/k$.
Interestingly enough, for  $ka \gg 1$, $f(ka) \approx ka$, for a discussion see Section I in METHODS.
Therefore, the limiting cross-section is $\approx 4a$, \emph{i.e.,} twice larger than the 
width of the obstacle. \cite{morse19861953methods} 
(This form of $\sigma_{fl}(k)$ is, of course, valid only when the energy of the phonon is much less than the pinning potential, \emph{i.e.,} $\omega(k) \ll \omega_0$. Relying on the existing experimental data \cite{he2013anomalous} 
we assume that the pinning potential $\omega_0$ is about few meV.)

Let us now touch upon a subtle question of the openness of the ensemble of the FPs which can be relevant for their localization. In the discussed model, the low-energy FPs cannot penetrate into the areas of strong contact with the substrate. Therefore, the most relevant channel connecting the FPs with the other degrees of freedom is the interaction with conducting electrons. The effect of this interaction can be estimated by comparing the amount of heat stored in the FPs with the rate of cooling of electrons. At low temperatures, most of the heat in the graphene layer is stored by the FPs, as they are the softest modes. The amount of this heat is $\propto T^2$. The rate of energy exchange between electrons and FPs, as discussed below in section "Thermal transport", is proportional to $T^3$. Therefore, the openness of the FPs, cannot effectively destroy their localization at low temperatures. 

\bigskip
\noindent
\textbf{Numerical study}

\noindent A general question has been addressed: 
If to compare with the electrons propagating in a disordered lattice, will the statistical properties of the eigemodes of the pinned elastic layer be the same or different? The question makes sense because for phonons in a pinned-suspended sheet there is no analogue of the on-site disordered potential $W$. Instead, there is concentration of the pinned sites. Furthermore, the FPs are described by the square of the Laplacian, rather than by the Laplacian in the case of electrons. To understand the general properties of the FPs in the presence of random pinning scatterers, we solve the equation of motion for the out-of-plane displacements using finite difference method on a $2d$ square lattice. We study a model  
in which the graphene sheet is completely attached at the pinning centers.
For that, we use 
discretized LHS of
Eq. \textbf{\ref{deq}} with condition $h=0$ at randomly chosen
pinned sites, so that one pinned site represents an attached area of the size $\approx 2a$ (See Fig. \ref{fig:wave} in METHODS for illustration of the model.)

\begin{figure}
\centering
\includegraphics[width=0.49\textwidth]{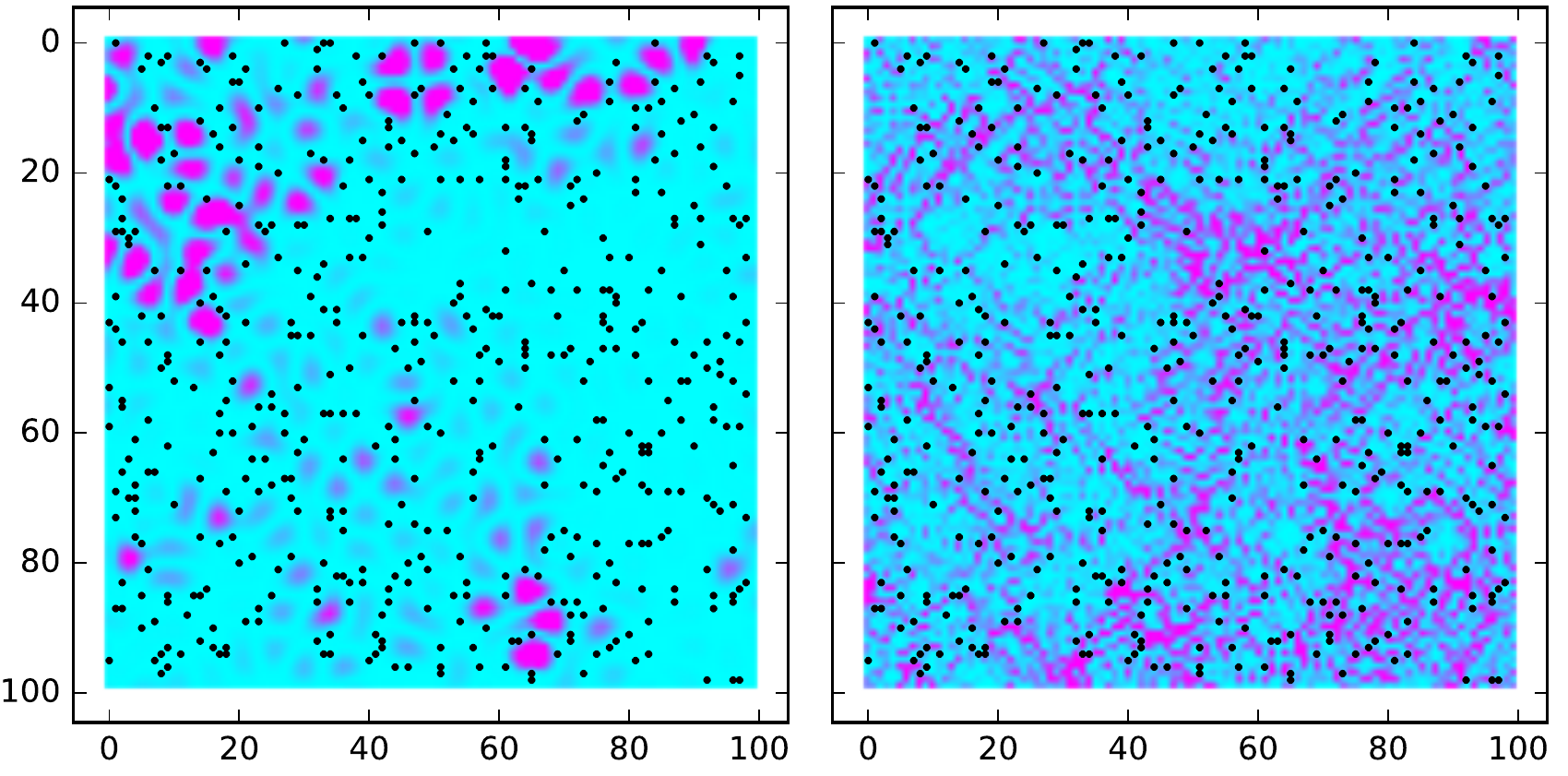}
\caption{The intensity of the phonon wavefunctions $\propto h^2$ for 5\% of the pinned sites at $E=0.5$ (left) and $E=3.1$ (right). Pinned sites are indicated as dots.}
\label{fig:wavefunction}
\end{figure}

We estimate the size of an attached area to be $2a\simeq 7$\emph{nm}. Typical distance between the pinning centers $a_i$ is around $20$\emph{nm}; we will assume that $n_i=a_i^{-2}$. The representative fraction of the pinned sites is $(2a/a_i)^2 \simeq(7/20)^2 \approx 12.5\%$. Correspondingly, we studied "samples" with $5\%-20\%$  of the 
pinned sites to determine statistical properties of the eigenmodes and eigenvalues. In doing so, we considered samples with periodic boundary conditions of the size 
up to 200$\times$200 sites. In what follows, we measure the energy eigenvalues $E$ in units of $\alpha/(2a)^2$ which approximately equals $0.08$K for the parameters mentioned above. We found out here that typical energy scale for \emph{strong} localization, $\omega_{loc}^{str}$, for 
the discussed concentration of pinned sites 
is a fraction of 1K. We believe that for graphene layer on the top of the SiO$_2$ substrate  $\omega_{loc}^{str}\approx 0.3$K is a realistic estimate. Eigenmodes at two representative energies are shown on Fig. \ref{fig:wavefunction} for a $200\times200$ sample (only $100\times100$ fragments are shown). 



\bigskip
\noindent
\textbf{Phononic "conductance"}

\noindent As is well known, localization is a quantum critical phenomenon. The peculiarity of $2d$ is that the critical point is at $1/g_{\square}=0$, where $g_{\square}$ is electrical conductance per square measured in units $e^2/(2\pi\hbar)$. At a finite $g_{\square}$, statistical properties are determined by the localization length $l_{loc}$, which is analogue of a correlation length at a quantum phase transition. A sample of size $L< l_{loc}(g_{\square})$ is in the regime of criticality which may take place in a very broad range of the sample sizes, because for  small $g_{\square}$ the localization length is exponentially large. \cite{abrahams1979scaling} 
A consequence of strong fluctuations of wave function amplitude in the critical region is the multifractality (that is when an eigenstate is extended but the occupied volume is noticeably smaller than the volume of the sample).
\begin{figure}
\centering
\includegraphics[width=0.5\textwidth]{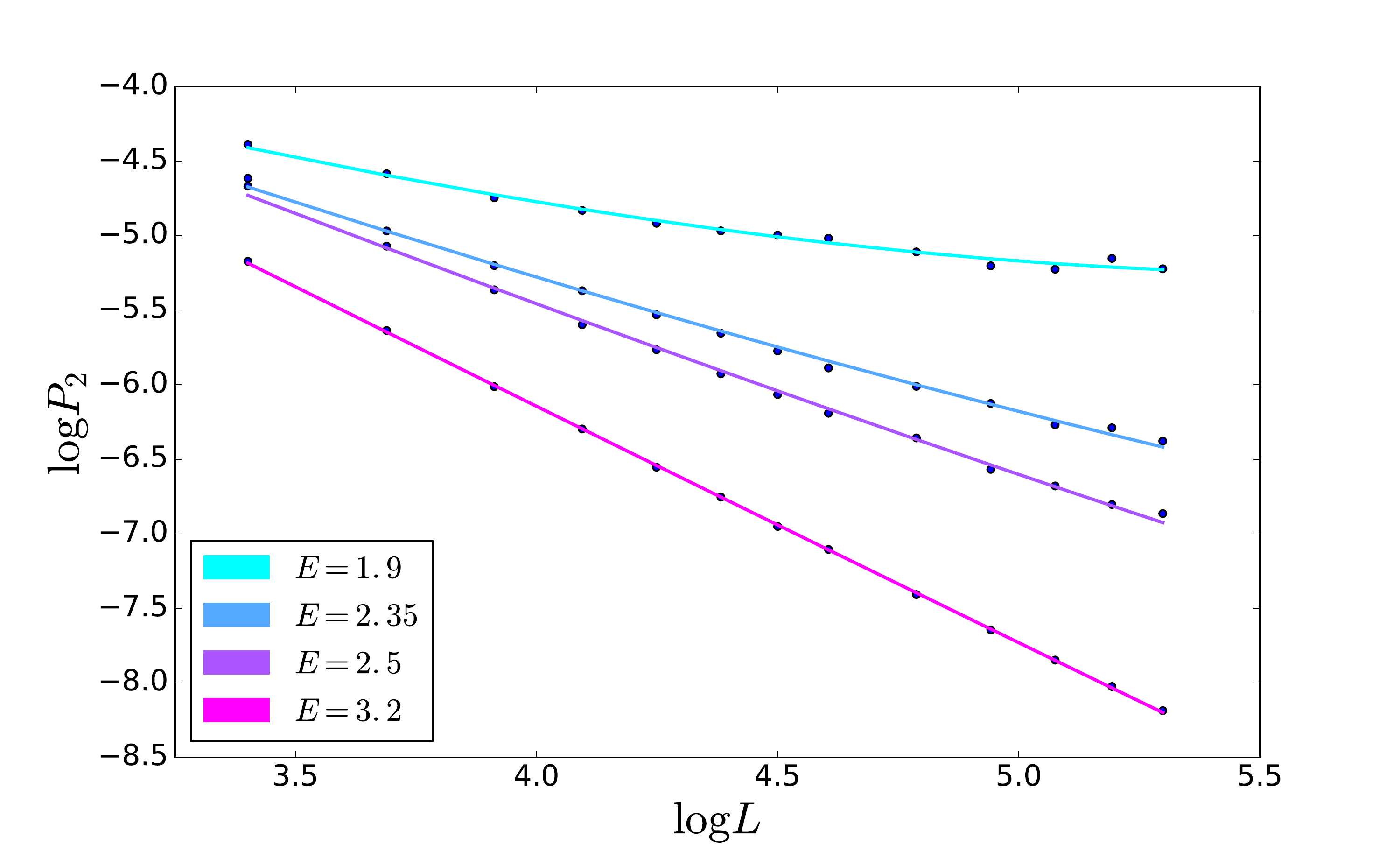}
\caption{Scaling of the IPR with the system size for 20\% of the pinned sites and several values of energy $E$. For the smallest energy the effect of the WL correction to $g_{ph}(L)$ is clearly seen.}
\label{fig:crossing}
\end{figure}
Turning back to disordered FPs, the first question that needs to be answered is: 
Is there
a transition to delocalized states at a certain energy (\emph{i.e.,} the "metal-insulator" transition with the mobility edge), or there is a crossover from strong to weak localization (WL)?

To figure this out, we 
first studied the dependence of the Inverse Participation Ratio (IPR) 
on the sample size $L$ for various phonon energies. A discussion of the IPR is given below in METHODS, section "Weak multifractality of eigenfunctions". 
(For more details the reader is referred 
to Ref. \onlinecite{mirlin2000statistics}.)
From our simulations, it is
clear that low-energy modes are localized: the IPR scales with the sample size to a finite value.
For higher energies, the behavior of the wave functions changes, see Fig. \ref{fig:wavefunction}, because the localization length  $l_{loc}$ starts to exceed the sample size. We are particularly interested in studying the FPs in this region when 
$l_{loc}\gg L$. Note 
that although the $2d$ Anderson model does not constitute a truly critical system, thanks to exponentially large (but still finite) localization length at $g_{\square}\gg 1$, the criticality takes place in a very broad range of the system sizes, $L\ll l_{loc}$. Therefore, $2d$ electrons at 
large $g_{\square}$ share many common properties with systems at the critical point of the metal-insulator transition. \cite{Abrahamsbook,evers2008anderson} 
As we shall see, similar physics 
holds also for our system of the FPs.

\emph{We proceed as follows:}
From the size dependence of the IPR at a given energy as shown on Fig. \ref{fig:crossing}, we extracted an energy-dependent fractal dimension. For disordered electronic system of a given symmetry class, the fractal dimension is determined by the conductance $g_{\square}$. For example, in the case of the Gaussian Orthogonal Ensemble (GOE), the size dependence of the IPR 
is described by the fractal dimension\cite{Wegner1980,Fal'koEfetovPRB52}
equal to $D_2(L)= 2-2/\pi g_{\square}(L)$, 
where dependence of $g_{\square}(L)$ on $L$ is due to the WL 
corrections. 
Thus, 
for each concentration of the pinned sites we can prescribe for different energies $\omega$ the corresponding value of the phonon "conductance" $g_{ph}(\%,\omega)$, using the expression for the fractal dimension $D_2$ for electrons. We defer the discussion of the dependence of $D_2$ on the sample size, $D_2(L)$, to
the end of section "Weak multifractality of eigenfunctions". 

For disordered electrons in $2d$, the well developed theory connects the behavior of various physical quantities with the value of the conductance. \cite{mirlin2000statistics} 
We have calculated numerically the same quantities for the FPs, 
using the values of  $g_{ph}$ extracted from IPR, and found a very good agreement with the theoretical predictions existing for the disordered electrons in the case of the GOE. Below we present 
some results of such calculations.

\begin{figure}
\centering
\includegraphics[width=0.5\textwidth]{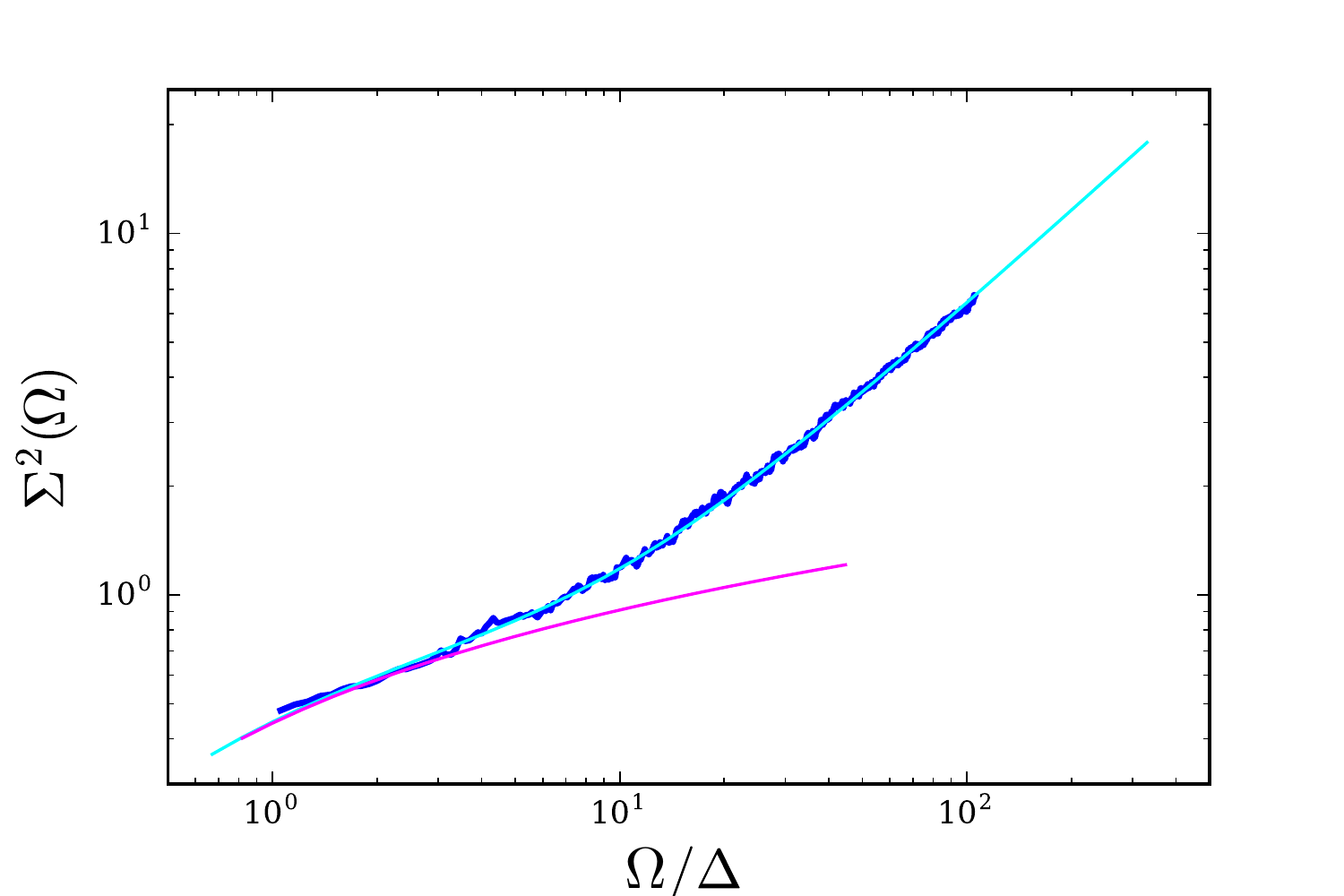}
\caption{Blue dots: level number variance $\Sigma^2(\Omega)$ in a sample $200\times200$ with 20\% of the pinned sites for $E\approx 3.5$. The theoretical fit with $g_{ph}=1.6$ is plotted by a solid line in cyan. The RMT result is given in magenta.}
\label{fig:variance}
\end{figure}

\bigskip
\noindent
\textbf{Statistical properties}

\noindent First, we have checked (see Fig. \ref{figPW} 
in METHODS, section "Energy level statistics") 
that the distribution function of the level spacing $P(s)$ for localized states has almost Poissonian statistics, while for metallic states it is of the Wigner-Dyson form. Next, in $2d$ it becomes especially interesting to study the 
variance $\Sigma^2(\omega,\Omega)$, which is a two-level correlation function characterizing the fluctuations of the number of levels $N$ in a strip of width $\Omega$ around the energy $\omega$: $\Sigma^2(\omega,\Omega)=\langle N^{2}(\Omega)\rangle-\langle N(\Omega)\rangle^{2}$. The reason why it is of particular interest is that, in contrast to $d=1$ and $3$, in two dimensions this quantity is directly related to the WL corrections. \cite {kravtsov1995level}  
Fig. \ref{fig:variance} demonstrates the level number variance as a function of the ratio $\Omega/\Delta$, where $\Delta$ is the average level spacing. It starts with the ergodic behavior described by the Random Matrix Theory (RMT). The ergodic regime holds up to $\Omega$ about the Thouless energy. For larger $\Omega$ there is a noticeable deviation: the variance starts to increase rapidly. The numerical results presented in Fig. \ref{fig:variance} are in full accord with the theoretical expression obtained by us for $d=2$; for details the reader is referred to section "Energy level statistics". 

To the best of our knowledge, this is the first demonstration of the mesoscopic fluctuations of the number of levels in $d=2$, while for the $3d$ Anderson model, the function $\Sigma^2(\omega,\Omega)$ was studied long ago. \cite{braun1995spectral}

\begin{figure}
\centering
\includegraphics[width=0.5\textwidth]{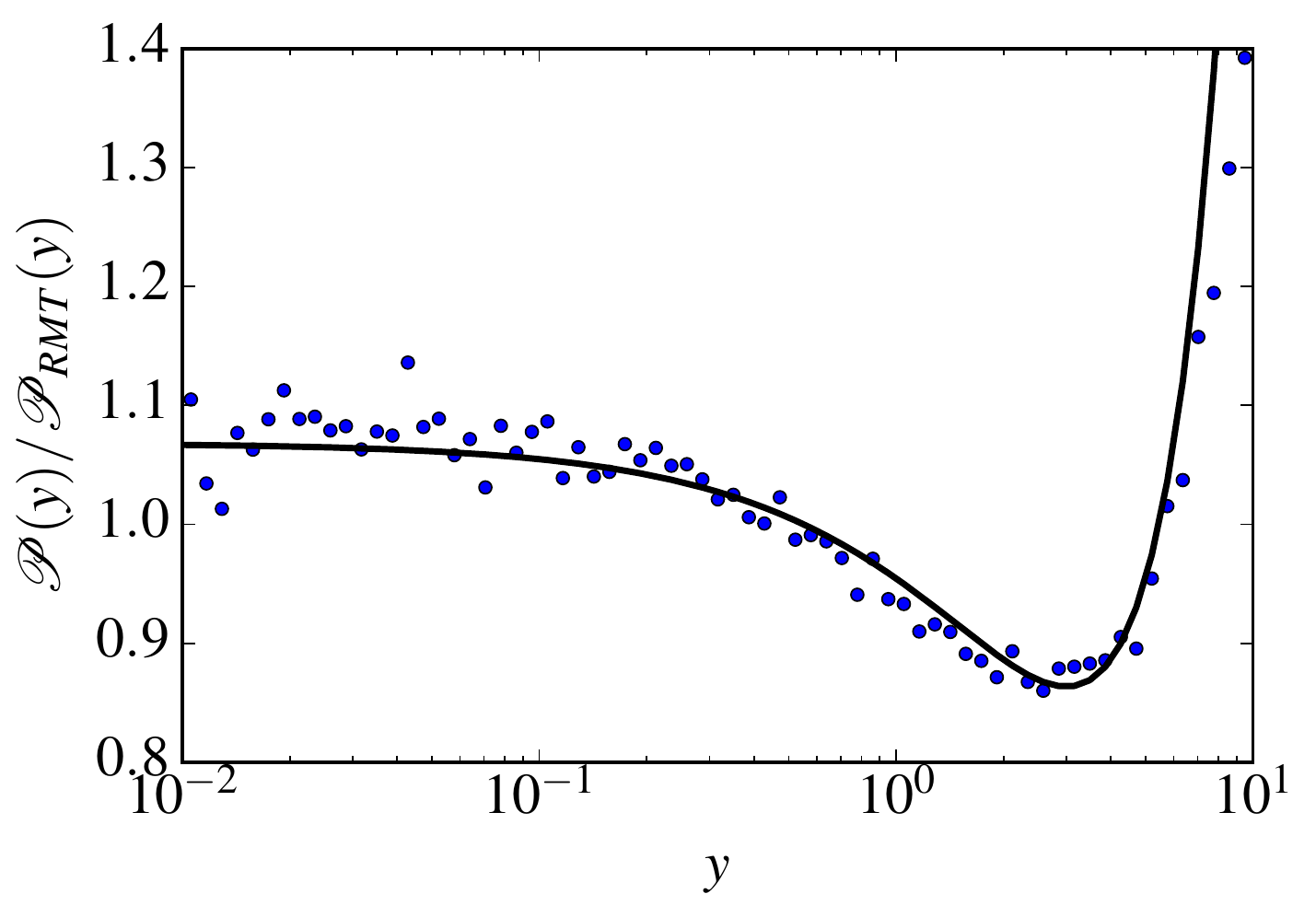}
\caption{Blue dots: the intensity distribution calibrated with respect to the RMT result for 10\% of the pinned sites at $E\approx 3.5$. Solid line: fit with theory (\onlinecite{fyodorov1995mesoscopic}) with $g=5.63$ and $L/l = 5$.}
\label{fig:fm}
\end{figure}

Another important statistical property is the 
distribution of the amplitudes of the eigenmodes, $\psi^2$, which is called the wave function intensity distribution $\mathcal{P}(y)$, where in our case $y$ is $\propto h^2$. 
For metallic granulas, in the ergodic regime described by the RMT, the intensity distribution is given 
by the Porter-Thomas distribution $\mathcal{P}_{RMT}(y)$. Owing to the fluctuations in the diffusive motion, there appear deviations from the ergodic behavior.
When calibrated with respect to $\mathcal{P}_{RMT}(y)$, the function $\mathcal{P}(y)$ yields a curve with a very specific non-monotonous shape. As Fig. \ref{fig:fm} shows, an excellent agreement with the theory of Ref. \onlinecite{fyodorov1995mesoscopic} 
is found. 

To summarize, we have calculated numerically a number of quantities characterizing statistical properties of FPs using the values for 
$g_{ph}$ extracted from the data for the IPR, and found a  very good agreement with the theoretical predictions existing for the disordered Anderson model electrons in the case of the Orthogonal Class of Universality. Furthermore, the theoretical expressions for the number of variance $\Sigma^2(\omega,\Omega)$ and the wave function intensity $\mathcal{P}(y)$, both are intimately connected with the effects of the WL originating from the Cooperons. The excellent agreement demonstrated in Figs. \ref{fig:variance} and \ref{fig:fm} justifies that in the discussed model the regime of WL is the same as in the Anderson model in $2d$. We believe that the reason for the observed universal behavior is that the FPs in the lattice with pinned sites are eventually described with the same Non-Linear $\sigma$-model as disordered electrons in the Orthogonal Class of Universality.

\bigskip

\noindent
\textbf{Estimate of the scales}

\noindent Let us estimate energy of the FPs at which a crossover from strong to WL occurs. Strong localization, $\omega \lesssim 1/\tau$, holds for momenta $k^2 \lesssim 8n_i$ that for our choice of $a_i$ yields 
$k<k_{loc}^{str}\approx0.14 nm^{-1}$. So far, we didn't consider the effect of strain. The strain $\bar u$, ignoring anisotropy, is known to add the term
$\rho \bar u v_L^2 k^2$ into the equation of motion, Eq. (\ref{deq}), where $v_L$ is velocity of the longitudinal phonons. In the isotropic approximation this yields $\omega (k)=\sqrt {(\alpha k^2)^2+ \bar u (v_L k)^2}$. 
One has to keep in mind that scattering of a FP by a high enough barrier doesn't depend on details, and the cross-section remains $4/k$, if $k < a^{-1}$. Then, for the linear spectrum the condition for strong localization is $k^2 \lesssim 4n_i$, which is similar to what we have got above. Typically, $\bar u$ is $\backsim 10^{-4}$, and the 
two terms in $\omega (k)$ are of comparable strength for the discussed scales.

So far, we have discussed point-like pinning centers only. In reality, the size of the attached areas can be comparable with the distances between them. Then,
owing to the factor $f\approx ka>1$, the energy of the strongly localized FPs can be a few times larger.
As we have already mentioned, a reasonable estimate for $\omega_{loc}^{str}$ for graphene layer on SiO$_2$ is $\approx0.3$K.
Furthermore, effects of the weak localization noticeably expand localization of the FPs. One may easily show that, as compared to the strong localization, the weak localization increases momenta of the FPs which undergo localization by a factor $\ln(L/l)$. (In this estimate, 
it is necessary to take into consideration the factor $f$ in the scattering cross-section.) 
Correspondingly, weak localization boosts the energy of the localized FPs 
by a factor $\ln^2(L/l)$. For a standard micron size sample, $\ln(\frac{L}{10\textrm{\emph{nm}}})\approx 5$. As a result, 
the energy of localized FPs may increase 
up to few K.
\bigskip

\noindent
\textbf{Effects of anharmonicity}

\noindent In writing Eq. (\ref{deq}) we have 
neglected effects of interaction of the FPs with the in-plane phonons (anharmonicity) and with ripples. \cite{NelsonPeliti1987,DavidGuitter1988,Doussal1992,Nelsonbook,LeDoussalRadzihovsky2017}
In graphene on a random substrate, FPs are not only scattered by the points of contact with the substrate, but they also excite acoustic phonons. One may check, following the calculations of Ref. \onlinecite{song2012} for disorder-assisted scattering, that at low temperatures the effect is vanishingly small.

Next, as it is well known anharmonicity yields a strong effect on the bending rigidity in graphene. The point is that the in-plane rigidity of graphene is extremely high
(i.e., it has a very large Young's modulus, $Y_0$), while a bending rigidity $\kappa_0$ is relatively modest. The anharmonicity transfers the strong in-plane rigidity into the bending one. The effect is controlled by the temperature, and is of the infrared origin. For small wave vectors $q\ll q_{th}$, the bending rigidity driven by thermal fluctuations is scale dependent: $\kappa_R(q)\sim \kappa_0{(q/q_{th})}^{-\eta}$ with the scaling exponent $\eta\approx0.8-0.85$.
\cite{los2009,bowick2017non-Hookean} The transition scale is given by $q_{th}=\sqrt{\frac{3k_BTY_0}{16\pi{\kappa_0}^2}}$. For graphene, at room temperature $q_{th}\approx1.6nm^{-1}$.
As a result, effective bending rigidity of an atomically thin graphene ribbon that are 10-100 micrometers in size at room temperature can be thousands times larger than at $T=0$.\cite{blees2015,nicholl2015,kosmrlj2016,los2016} (At $T=0$, the
quantum non-linear effects lead to only logarithmic corrections, hence generally much smaller than
the power-law renormalization produced by thermal fluctuations.\cite{KatsLebedev2014})

Strong effects caused by the renormalizations, which have been mentioned above, correspond to the vanishing wave vector $q\rightarrow0$. However, above the transition vector $q_{th}$ thermal fluctuations caused by anharmonicity are no longer significant and $\kappa_R(q)\approx \kappa_0$. 
In graphene, the energy of a FP with the wave vector $q_{th}$ is equal to $\omega_{th}\approx 0.03k_BT\ll k_BT$. Thus, there is a substantial energy gap between the thermal phonons and phonons, for which the effects of the anharmonicity are relevant.
For our estimate of $\omega_{loc}^{str}\approx 0.3$K, temperature should be about 10K or higher to influence 
our analysis of statistical properties of the FPs on the SiO$_2$ substrate.
Furthermore, in our analysis, 
we were mostly interested in FPs with the frequency exceeding $\omega_{loc}^{str}$, so that they can propagate between pinned regions colliding randomly with them.

\bigskip
\noindent \textbf{DISCUSSION}

\noindent We shall discuss now the implications of the FPs on the thermal transport of a graphene layer placed on a corrugated SiO$_2$ substrate. We argue that traces of localization 
of the FPs may have been observed in the experiments at low temperatures.

\bigskip
\noindent
\textbf{Thermal transport}

\noindent The temperature behavior of overheating in graphene on the top of SiO$_2$ at low temperatures (see Refs. \onlinecite{borzenets2013phononbottleneck,KinChungFong2013fig3,prober2016,bistritzer2009,song2012,chen2012})
has not been fully
understood, yet. Theoretically, the heat flux from electrons to the FPs is known \cite{song2012}  
to be $\propto T_{el}^3$ (correspondingly, the reversed flux from the FPs to electrons is $\propto T_{ph}^3$). 
We consider the heat exchange of electrons with the FPs as realistic explanation for the total power $P\propto T^\delta$ with the exponent $\delta=3$ observed \cite{borzenets2013phononbottleneck, KinChungFong2013fig3} 
at low temperatures and \emph{far away} from the neutral point. At concentrations of the electric carriers, electrons or holes, $n\sim 10^{12} cm^{-2}$ the alternative explanation of $\delta=3$ with the use of the result 
obtained for the case of the \emph{unscreened} deformation potential \cite{chen2012} is not realistic; a discussion of this point is given below in section "Electron-phonon interaction at low temperatures". 

Moreover, we believe that the existence of localized FPs may explain the experimental result  of Ref. \onlinecite{KinChungFong2013fig3} for cooling rate of graphene at the lowest temperatures $T\lesssim0.85$K.
In this regime, the cooling is dominated 
by the electron heat diffusion along the sample.
However, the Lorenz number $\mathcal{L}$ estimated in this way, was found to be 35\% above  its nominal value $\mathcal{L}_0$.
It has been shown in Refs. \onlinecite{Schwiete2014,Schwiete2016a,Schwiete2016b} 
that neither the Fermi liquid nor renormalization-group corrections in disordered $2d$ 
electron systems 
can modify the Lorenz number $\mathcal{L}$ 
and, therefore, the result requests for an explanation.
Here we argue that the heat 
exchange of electrons with the localized FPs, $\delta P_{ep}^{loc}$, may resolve this problem.
The point is that the heat exchange with the localized phonons 
has the form \emph{imitating} the electron heat diffusion contribution, $\delta P_{\textrm{ep}}^{\textrm{loc}}\propto T (T_{el}-T_{ph})$. 

Let us comment upon $\delta P_{ep}^{loc}$. First of all, we recall that interaction of an electron with the FPs is described by the two-phonon processes. Correspondingly, the heat exchange between the electrons and FPs, which is proportional to square of the two-phonon amplitude,
contains $\emph{four}$ powers of the FP-momenta. However, the localized FPs are not goldstone modes anymore. For localized FPs, the momenta that enter into the matrix elements of the electron-FP interaction should be substituted by the inverse of the localization length. As a result, two powers of frequencies 
in the expression for the heat flux $P$ saturate at $\omega \simeq \omega_{loc}^{str}$. This, however, leads to a dramatic consequences. The factor describing the dependence on the occupation numbers in the case of the two-phonon processes for temperatures larger than the energy of the phonons diverges like $\omega^{-2}$. The diverging integration should be cut-off at energies typical for the localized FPs.
As a result one gets a contribution to the cooling rate of the order $\delta P_{\textrm{ep}}^{\textrm{loc}}\propto \omega_{loc}^{str} T (T_{el}-T_{ph})$. The obtained correction to the heat flux has just the form of the electron heat diffusion. In order to obtain experimentally observed magnitude of the deviation of $\mathcal{L}$ from $\mathcal{L}_0$, one has to suggest $\omega_{loc}^{str} \sim 0.3$K. This is in full correspondence with our expectations of the scale energies where localization of the FPs takes place.


\bigskip
\noindent
\textbf{Summary}

\noindent Studying effects of disorder on the properties of elastic membranes has a long history.\cite{Radzihovsky1991,Morse1992,leDoussal1993,gornyi12,kosmrlj2013,gornyi2015rippling,Lopez-Polin2015,LeDoussalRadzihovsky2017} However, a layer placed on the top of a corrugated substrate, which has been discussed in the present work, is very different from
the disordered membranes considered so far. First, disorder here is external rather than internally quenched. Next, 
disorder pins rigidly the height of randomly chosen points of the layer, rather than acting on the metric and curvature tensors describing the deformation of the membrane. Questions of localization of the FPs to the best of our knowledge have not been addressed previosly.

Graphene layers on top of SiO$_2$ substrates are expected to play an important role in applications related to thermal transport and for ultrasensitive bolometry. While the graphene sheet is pinned at random points, the FPs 
may exist in between due to the corrugation typical for this substrate surface. We have argued that in this system Anderson localization of low-energy FPs 
develop. We showed that the ensemble of flexural phonons in a pinned-suspended flexible sheet is statistically identical to an ensemble of disordered electrons, despite the very different underlying mathematical descriptions. 
Localization of flexural phonons should be important for thermal transport in such hybrid systems even at not very low temperatures.
Traces of localization 
of the FPs may have been already observed in the experiments. 
Let us note that we have already shown that FPs give a significant contribution to dephasing rate of electrons in graphene. \cite{tikhonov2014dephasing}
Here, we have argued that localization of the FPs opens interesting perspectives for thermal transport.

\clearpage
\onecolumngrid
\begin{center}
\large{\textbf{METHODS}}
\end{center}
\setcounter{section}{0}

\section{Scattering of a flexural phonon by a rigid obstacle}
\begin{figure}[H]
\centering
\includegraphics[width=0.33\textwidth]{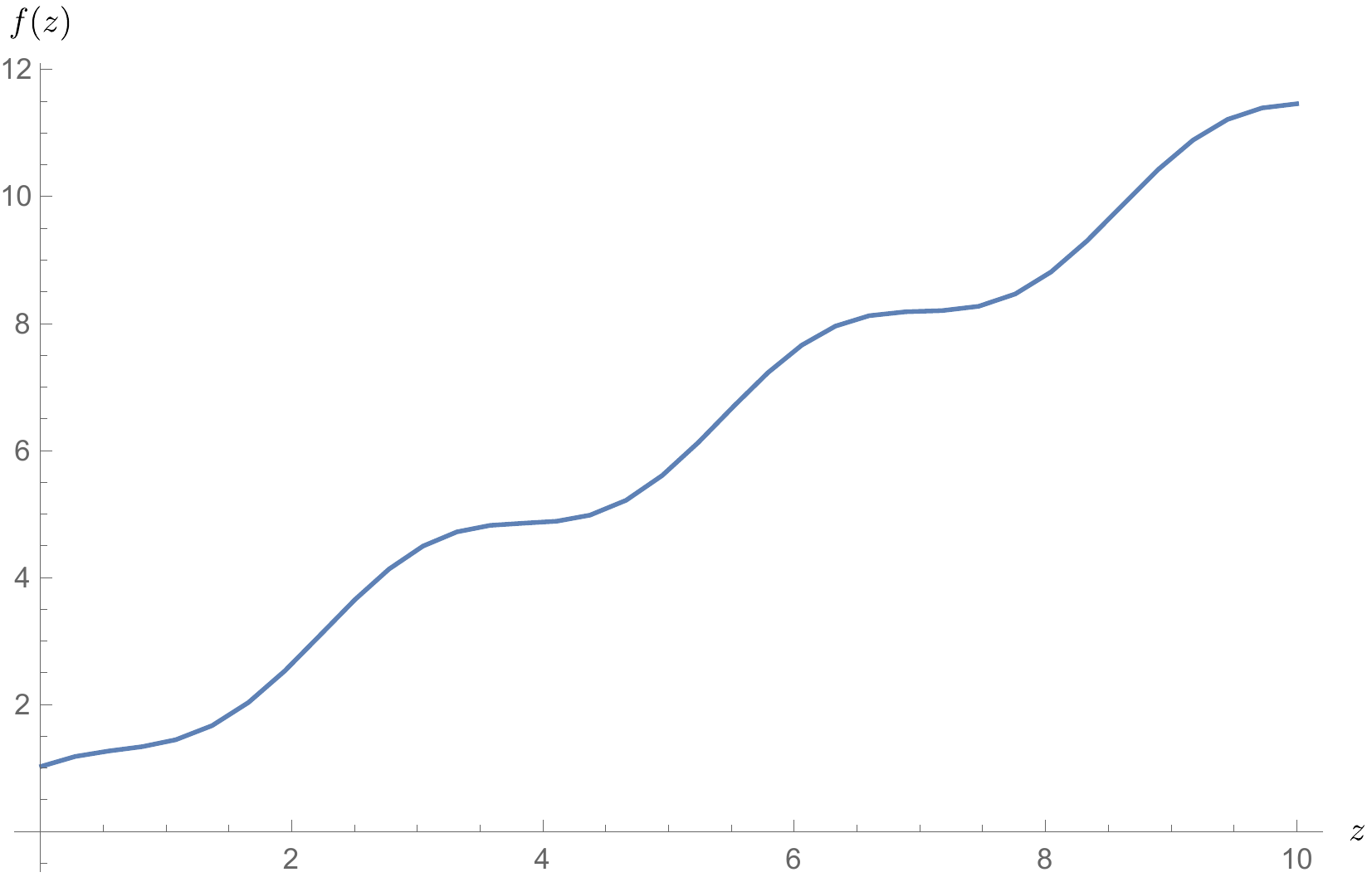}
\caption{Function $f(z)$, determining the scattering cross-section, $\sigma_{fl}=\frac4k f(ka)$.}
\label{figcs}
\end{figure}

According to Ref. \onlinecite{norris1995scattering}, the scattering cross-section of the FP by a rigid obstacle of the radius $a$ equals $\sigma_{fl}=\frac{4}{k}f(ka)$ where $f(z)$ is given by the following expression:
\begin{equation}
f(z)=\textrm{Re}\sum_{n=0}^{\infty}\varepsilon_n\frac{J_n(z) K'_n(z) - J'_n(z) K_n(z)}{H^{(1)}_n(z) K'_n(z) - H^{(1)\prime}_n(z) K_n(z)},
\end{equation}
where $\varepsilon_0=1,\;\varepsilon_{n>0}=2$. The function $f(z)$ with asymptotes $f(0)=1$ and $f(z\gg 1)\approx z$ is shown in Fig. \ref{figcs}. For short wave-lengths the limiting cross-section is twice larger than the width of the obstacle.
Note that the factor $f$ is important for the effectiveness of the weak localization in samples of large size.

\section{Weak multifractality of eigenfunctions}

The spatial distribution of wave functions is conveniently characterized by inverse participation ratios:
\begin{equation}
P_{q}=\int d^{d}r|\psi(r)|^{2q}.
\end{equation}
After sample average, $\langle P_{q}\rangle$ shows the scaling behavior with the system size $L$:
\begin{equation}
\langle P_{q}\rangle\sim L^{-D_q(q-1)}.\label{multifr2}
\end{equation}
Obviously, in the insulating state $D_{q}=0$, while in a metal $D_{q}=d$.
At a critical point, $D_{q}$ is a fractional which leads to anomalous scaling behavior in $\langle P_{q}\rangle$. This is a manifestation of the wave function multifractality,
which is a consequence of the spatial correlations of the wave function. 

In $2d$, the inverse participation ratios scale as
\begin{equation}
\langle P_{q}\rangle \simeq (2q-1)!!L^{-2(q-1)} \left(\frac{L}{l}\right)^{\frac{1}{\pi g}q(q-1)},\label{multifr4}
\end{equation}
that corresponds to 
\begin{equation}
D_{q}=2-\frac{q}{\pi g}.\label{multifr5}
\end{equation}
Here, the deviation of $D_{q}$ from dimension $2$ is determined by a small parameter $1/\pi g$.
The above result \textbf{\ref{multifr5}} was first obtained by Wegner \cite{Wegner1980} via the renormalization group calculations for a system of disordered (non-interacting) electrons; see also  Ref. \onlinecite{Fal'koEfetovPRB52}. The dimensionless parameter $g$ is a conductance of a sample measured in quantum units. In analogous to weak localization, the phenomena is coined ''weak multifractality''.

\begin{figure}[H]
\centering
\includegraphics[width=0.5\textwidth]{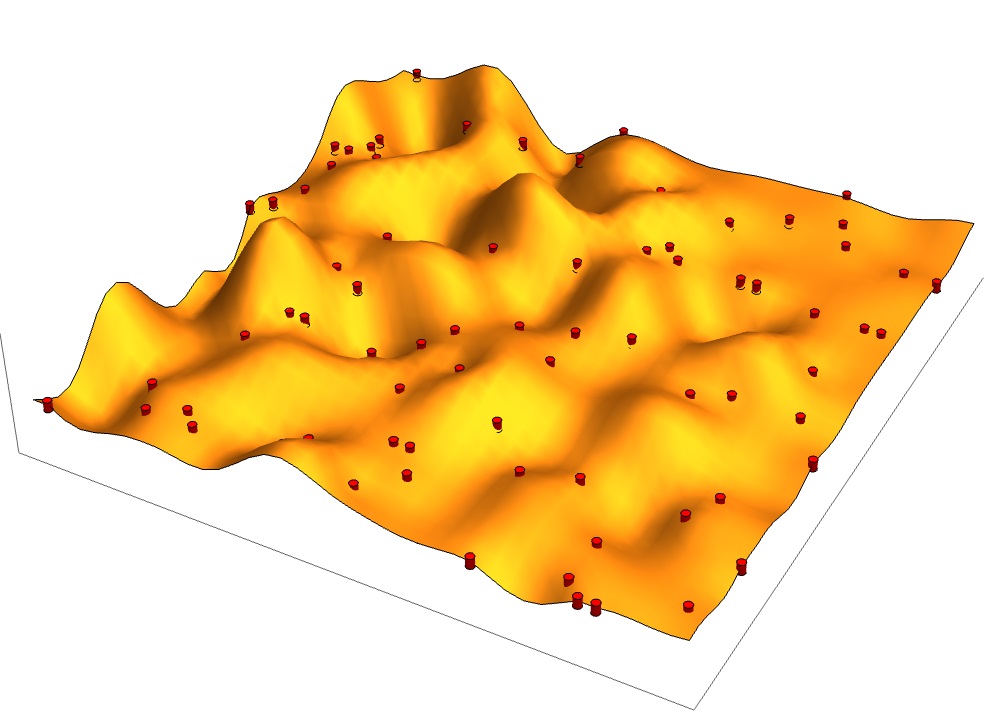}
\caption{Elastic flexible $2d$ sheet on a substrate: pinning centers are indicated by red cylinders.}
\label{fig:wave}
\end{figure}

We study a model of random pinning centers,
in which the graphene sheet is completely attached at the pinning centers; see 
Fig. \ref{fig:wave}. In our 
analysis of multifractality of the FPs 
we used $q=2$. The corresponding inverse participation ratio, $P_{q=2}$, was denoted as IPR. This quantity allowed us to extract the phononic "conductance", which we used for the statistical analysis of our system. Note that, because of the absence of the genuine critical point in $2d$, $g$ becomes size-dependent, \emph{i.e.,} $g(L)=g_0-(1/\pi)  \ln L/l$ where $l$ is the mean-free path of FPs at a given energy, and $g_0=g(l)$. This implies that for each scale $L$ one can use the standard formula given above, but with slowly varying $g(L)$ in the exponent. 
This is possible because corrections to $g$ are 
not large in a finite size sample, and owing to the slow dependence of $g$ on spatial scale $L$. With this procedure, we have obtained an excellent agreement between the theory of the logarithmic corrections to the conductivity and our numerical results as it is shown in Fig. 2 of the main text. 

\section{Energy Level Statistics}
Several quantities are introduced to measure the fluctuations of energy
levels $\omega_{n}$, such as the distribution function of level spacing $P(s)$, and the level number variance $\Sigma^{2}(\omega, \Omega)$.

\begin{figure}
\centering
\includegraphics[width=0.5\textwidth]{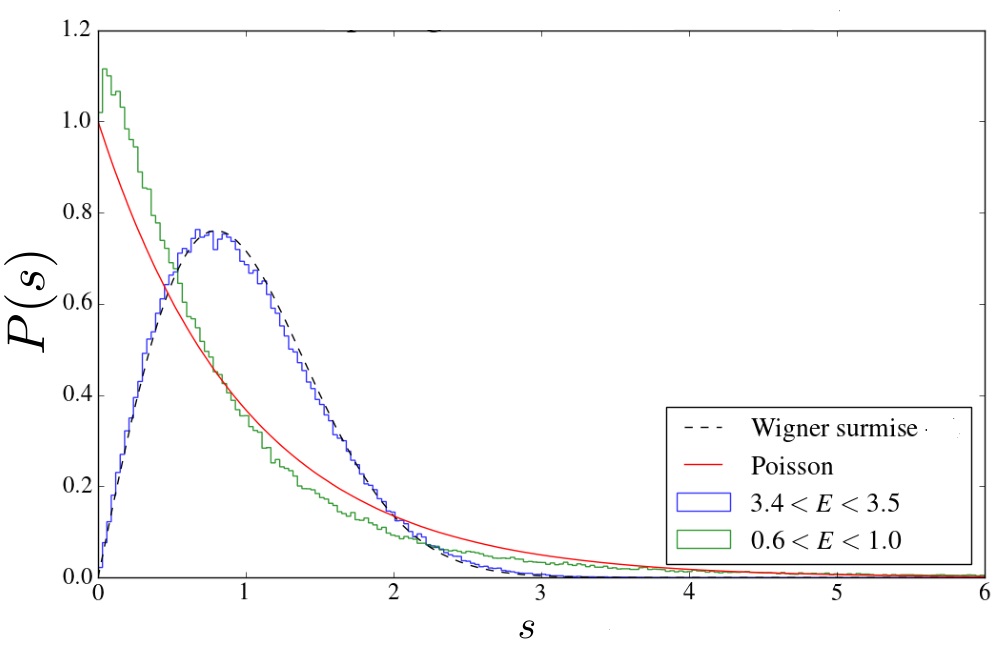}
\caption{Crossover from the Poisson to Wigner-Dyson level statistics at 20\% disorder.}
\label{figPW}
\end{figure}

Random Matrix Theory (RMT) could be used to describe these quantities in ergodic systems (e.g., for electrons in metallic granules). Here, we will focus on the case of Gaussian Orthogonal Ensemble (GOE). Then the distribution function of level spacing is well described by the Wigner surmise: $P_{o}(s)=\frac{\pi}{2}s\exp(-\frac{\pi}{4}s^{2})$, where $s=\frac{\omega_{n+1}-\omega_n}{\Delta}$ and $\Delta$ is mean level-spacing.
In the localized phase the level correlations are absent, and the distribution function of level spacings is Poissonian: $P(s)=\exp(-s)$. In our system of the FPs the crossover from the localized to delocalized behavior is illustrated in Fig. \ref{figPW}.

In the RMT, the level number variance $\Sigma^{2}(\omega, \Omega)$ increases logarithmically with $\Omega$.
For $\Omega\gg\Delta$, it varies as $\Sigma^{2}(\Omega)=\frac{2}{\pi^{2}}\ln(2\pi\langle N\rangle)+c_{\beta}+O(\langle N\rangle)$,
where $\langle N\rangle=\Omega/\Delta$, and $c_{\beta}$ is a known constant. This, however, is far from the true 
behavior (as one can see in Fig. 3 of the main text). 
For a further analysis it useful that
function $\Sigma^{2}(\omega, \Omega)$ is closely related with the two-level correlation function $R(\Omega)$, which
is defined as
\begin{equation}
R(\Omega)=\frac{\left\langle \rho(\omega-\Omega/2)\rho(\omega+\Omega/2)\right\rangle }{\left\langle \rho(\omega)\right\rangle ^{2}}-1.\label{eq:DOScorr}
\end{equation}
Here $\rho(\omega)=V^{-1}\sum_{n}\delta(\omega-\omega_{n})$ is the Density of States (DOS),
and $\left\langle \rho\right\rangle $
is the average DOS, 
which is related to the mean level-spacing $\Delta$ as $\left\langle \rho\right\rangle =\frac{1}{\Delta V}$.
The connection between the two correlation functions can be presented 
as
\begin{eqnarray}
\Sigma^{2}(\omega,\Omega) = 2\int_{0}^{\langle N\rangle}(\langle N\rangle-s)R(s)ds, \label{sigmaR}
\end{eqnarray}
or, equivalently, $R(\Omega)=\frac{\Delta^{2}}{2}\frac{\partial^{2}\Sigma^{2}(\Omega)}{\partial\Omega^{2}}$.

Kravtsov and Lerner recognized in Ref. \onlinecite{kravtsov1995level} that in $2d$, unlike $d=1$ and $d=3$,
the level correlation function $R\left(\Omega\right)$ and, hence, $\Sigma^{2}(\omega, \Omega)$ are governed entirely by the weak localization corrections. 
They found that $R(\Omega)=\frac{\Delta}{\pi^{2}}\sum_{q}\mbox{Re}\frac{1}{\left(Dq^{2}-i\Omega\right)^{2}}$ with the diffusion constant $D\to D+\delta D$, where
$\delta D=\mbox{\ensuremath{-D\frac{\Delta}{\pi}\mbox{\ensuremath{\sum_{Q}}}\frac{1}{DQ^{2}-i\Omega}}}$.
This gives
\begin{align}
R(\Omega) & =\frac{\Delta}{\pi^{2}}\textrm{Re}\sum_{q}\frac{1}{\left[Dq^{2}\left(1-\frac{\Delta}{\pi}\sum_{Q}\frac{DQ^{2}}{\left(DQ^{2}\right)^{2}+\Omega^{2}}\right)-i\Omega\left(1+\frac{\Delta}{\pi}\sum_{Q}\frac{Dq^{2}}{\left(DQ^{2}\right)^{2}+\Omega^{2}}\right)\right]^{2}}\\
 & \approx\frac{\Delta}{\pi^{2}}\textrm{Re}\sum_{q}\frac{1}{\left[Dq^{2}-i\Omega\left(1+\frac{\Delta}{\pi}\sum_{Q}\frac{Dq^{2}}{\left(DQ^{2}\right)^{2}+\Omega^{2}}\right)\right]^{2}}.
\end{align}
Now, instead of splitting this expression into two parts as it was done in Ref. \onlinecite{kravtsov1995level}, we use Eq. \textbf{\ref{sigmaR}} to calculate the level number
variance. After integration in $Q$ and $s$, this yields
\begin{equation}
\Sigma^{2}(\langle N\rangle)=\Sigma_{RMT}^{2}(\langle N\rangle)+\frac{1}{\pi^{2}}\sum_{n\neq0}\left[\log\left(1+\frac{\pi\langle N\rangle/g_{*}^{2}}{n^{2}}+\frac{\langle N\rangle^{2}/g_{*}^{2}}{n^{4}}\right)-\frac{\pi\langle N\rangle/g_{*}^{2}}{n^{2}}\right],\label{eq:Zhao}
\end{equation}
where $g_{*}=2\pi g+\frac{\pi^{2}}{4}$. 
We use the expression determined by formula \textbf{\ref{eq:Zhao}} for fitting the numerical data presented in 
Fig. 3 in the main text.

\section{Statistical properties of the wave functions}

Porter and Thomas were first who studied the distribution of eigenfunction amplitudes within the RMT framework.
Their result shows simply Gaussian distribution, which leads to the following distribution of the intensities $y$:
\begin{eqnarray}
{\cal P}_{RMT}(y) & =\frac{e^{-y/2}}{\sqrt{2\pi y}},\label{e3.2}
\end{eqnarray}
Here $y_{i}=V|\psi_{i}^{2}|$ is normalized in such a way that
$\langle y\rangle=1$.

The supersymmetric field theory was applied to the study of the eigenfunction statistics in a $d$-dimensional
disordered system. The eigenfunction intensity $y=V|\psi^{2}(r_{0})|$ in a point $r_{0}$
is distributed as:
\begin{equation}
{\cal P}(y)=\frac{1}{\nu}\left\langle \sum_{\alpha}\delta(V|\psi_{\alpha}(r_{0})|^{2}-y)\delta(E-E_{\alpha})\right\rangle.\label{anom9}
\end{equation}
For $d=2$, and not too large $y$, one can calculate perturbatively the deviations from the RMT distribution ${\cal P}(y)$. The corrected distribution function was found by Fyodorov and Mirlin in Ref. {\onlinecite{fyodorov1995mesoscopic}:
\begin{equation}
{\cal P}(y)\approx{\cal P}_{\textrm{RMT}}(y)\left[1+\frac{\kappa}{2}\left(\frac{3}{2}-3y+\frac{y^{2}}{2}\right)\right]\label{eq:dist-small O}
\end{equation}
with $\kappa = \frac{1}{\pi g}\ln\frac{L}{l}$. Notice that the expression in the square brackets is non-monotonous. A peculiar behavior of this expression as a function $y$ can be seen in Fig. 4 of the main text.

\section{Electron-phonon interaction at low temperatures}

Some recent experiments interpreted 
the cooling rate in terms of the weakly screened electron-phonon (e-ph) interaction down to very low temperatures.
In particular, 
in Ref. 
\onlinecite{KinChungFong2013fig3} the $T^3$ law has been observed down to
$0.5$K for sample D3. Such a behaivior 
corresponds to the dirty  
regime with negligible screening of the e-ph deformation potential according to the Table I presented in Ref. \onlinecite{chen2012}. We believe, however, that applicability of the screenless approximation for the e-ph interaction at such low temperatures and concentration of carriers $n\sim 10^{12}cm^{-2}$ is highly questionable.

To illustrate our point, let us estimate, following Ref.~
\onlinecite{chen2012}, temperature \emph{above} which screening becomes irrelevant in the dirty regime $T<T_{\textrm{dis}}=\hbar s/lk_B$. Comparing the expressions for the energy flux 
at strong and weak screening \cite{chen2012}, one finds that crossover from the $T^5$ to the observed $T^3$ behavior should happen at $T>T_*$ which satisfies 
\begin{equation}
0.1\kappa^2\left(\frac{T_*}{T_{BG}}\right)^2\approx 1,
\end{equation}
with $T_{BG}=54\sqrt{n/10^{12}\textrm{cm}^{-2}}$ for the Bloch-Gruneisen temperature and $\kappa$ for an effective dielectric constant (for graphene on SiO$_2$ substrate, $\kappa\approx 3$). Here we expressed the result in terms of $T_{BG}$ using the fact that in graphene the inverse of the screening radius is of the order of the Fermi momentum. The unscreened $T^3$ behavior may occur at temperatures higher than $T_*$ assuming that it is less than $T_{\textrm{dis}}$. (For samples D1 and D3 in Ref. \onlinecite{KinChungFong2013fig3} one has $T_{\textrm{dis}}\approx 40$K.)

For the aforementioned sample D3, $T_{BG}\approx 80$K. Thus, one may expect for the crossover temperature $T_*\approx\frac{250}{\kappa}\approx 80$K. It is clear that pushing $T_*$ down to $1$K region for the discussed densities of charge carriers would require an unrealistic value of $\kappa$. We, therefore, believe that this mechanism may 
be discarded for explanation of the $T^3$ scaling of the electron-phonon heat flux observed at low temperatures.

\bibliography{flexural}

\begin{thebibliography}{60}
\expandafter\ifx\csname natexlab\endcsname\relax\def\natexlab#1{#1}\fi
\expandafter\ifx\csname bibnamefont\endcsname\relax
  \def\bibnamefont#1{#1}\fi
\expandafter\ifx\csname bibfnamefont\endcsname\relax
  \def\bibfnamefont#1{#1}\fi
\expandafter\ifx\csname citenamefont\endcsname\relax
  \def\citenamefont#1{#1}\fi
\expandafter\ifx\csname url\endcsname\relax
  \def\url#1{\texttt{#1}}\fi
\expandafter\ifx\csname urlprefix\endcsname\relax\def\urlprefix{URL }\fi
\providecommand{\bibinfo}[2]{#2}
\providecommand{\eprint}[2][]{\url{#2}}

\bibitem[{\citenamefont{Geim and Grigorieva}(2013)}]{GeimGrigorieva}
\bibinfo{author}{\bibfnamefont{A.}~\bibnamefont{Geim}} \bibnamefont{and}
  \bibinfo{author}{\bibfnamefont{I.}~\bibnamefont{Grigorieva}},
  \bibinfo{journal}{Nature} \textbf{\bibinfo{volume}{499}},
  \bibinfo{pages}{419} (\bibinfo{year}{2013}).

\bibitem[{\citenamefont{Abrahams}(2010)}]{Abrahamsbook}
\bibinfo{author}{\bibfnamefont{E.}~\bibnamefont{Abrahams}},
  \emph{\bibinfo{title}{50 years of Anderson Localization}}
  (\bibinfo{publisher}{World Scientific}, \bibinfo{year}{2010}).

\bibitem[{\citenamefont{Morse and Feshbach}(1986)}]{morse19861953methods}
\bibinfo{author}{\bibfnamefont{P.}~\bibnamefont{Morse}} \bibnamefont{and}
  \bibinfo{author}{\bibfnamefont{H.}~\bibnamefont{Feshbach}},
  \bibinfo{journal}{NewYork: MeGraw-Hill}  (\bibinfo{year}{1986}).

\bibitem[{\citenamefont{Rayleigh}(1896)}]{rayleigh1896theory}
\bibinfo{author}{\bibfnamefont{J.~W. S.~B.} \bibnamefont{Rayleigh}},
  \emph{\bibinfo{title}{The theory of sound}}, vol.~\bibinfo{volume}{2}
  (\bibinfo{publisher}{Macmillan, London}, \bibinfo{year}{1896}).

\bibitem[{\citenamefont{Ishigami et~al.}(2007)\citenamefont{Ishigami, Chen,
  Cullen, Fuhrer, and Williams}}]{ishigami2007nanoletters}
\bibinfo{author}{\bibfnamefont{M.}~\bibnamefont{Ishigami}},
  \bibinfo{author}{\bibfnamefont{J.}~\bibnamefont{Chen}},
  \bibinfo{author}{\bibfnamefont{W.}~\bibnamefont{Cullen}},
  \bibinfo{author}{\bibfnamefont{M.}~\bibnamefont{Fuhrer}}, \bibnamefont{and}
  \bibinfo{author}{\bibfnamefont{E.}~\bibnamefont{Williams}},
  \bibinfo{journal}{Nano Letters} \textbf{\bibinfo{volume}{7}},
  \bibinfo{pages}{1643} (\bibinfo{year}{2007}).

\bibitem[{\citenamefont{Stolyarova et~al.}(2007)\citenamefont{Stolyarova, Rim,
  Ryu, Maultzsch, Kim, Brus, Heinz, Hybertsen, and Flynn}}]{stolyarova2007PNAS}
\bibinfo{author}{\bibfnamefont{E.}~\bibnamefont{Stolyarova}},
  \bibinfo{author}{\bibfnamefont{K.~T.} \bibnamefont{Rim}},
  \bibinfo{author}{\bibfnamefont{S.}~\bibnamefont{Ryu}},
  \bibinfo{author}{\bibfnamefont{J.}~\bibnamefont{Maultzsch}},
  \bibinfo{author}{\bibfnamefont{P.}~\bibnamefont{Kim}},
  \bibinfo{author}{\bibfnamefont{L.~E.} \bibnamefont{Brus}},
  \bibinfo{author}{\bibfnamefont{T.~F.} \bibnamefont{Heinz}},
  \bibinfo{author}{\bibfnamefont{M.~S.} \bibnamefont{Hybertsen}},
  \bibnamefont{and} \bibinfo{author}{\bibfnamefont{G.~W.} \bibnamefont{Flynn}},
  \bibinfo{journal}{Proceedings of the National Academy of Sciences}
  \textbf{\bibinfo{volume}{104}}, \bibinfo{pages}{9209} (\bibinfo{year}{2007}).

\bibitem[{\citenamefont{Geringer et~al.}(2009)\citenamefont{Geringer, Liebmann,
  Echtermeyer, Runte, Schmidt, R{\"u}ckamp, Lemme, and
  Morgenstern}}]{geringer2009intrinsic}
\bibinfo{author}{\bibfnamefont{V.}~\bibnamefont{Geringer}},
  \bibinfo{author}{\bibfnamefont{M.}~\bibnamefont{Liebmann}},
  \bibinfo{author}{\bibfnamefont{T.}~\bibnamefont{Echtermeyer}},
  \bibinfo{author}{\bibfnamefont{S.}~\bibnamefont{Runte}},
  \bibinfo{author}{\bibfnamefont{M.}~\bibnamefont{Schmidt}},
  \bibinfo{author}{\bibfnamefont{R.}~\bibnamefont{R{\"u}ckamp}},
  \bibinfo{author}{\bibfnamefont{M.~C.} \bibnamefont{Lemme}}, \bibnamefont{and}
  \bibinfo{author}{\bibfnamefont{M.}~\bibnamefont{Morgenstern}},
  \bibinfo{journal}{Physical Review Letters} \textbf{\bibinfo{volume}{102}},
  \bibinfo{pages}{076102} (\bibinfo{year}{2009}).

\bibitem[{\citenamefont{Deshpande et~al.}(2009)\citenamefont{Deshpande, Bao,
  Miao, Lau, and LeRoy}}]{leroy2009spatially}
\bibinfo{author}{\bibfnamefont{A.}~\bibnamefont{Deshpande}},
  \bibinfo{author}{\bibfnamefont{W.}~\bibnamefont{Bao}},
  \bibinfo{author}{\bibfnamefont{F.}~\bibnamefont{Miao}},
  \bibinfo{author}{\bibfnamefont{C.~N.} \bibnamefont{Lau}}, \bibnamefont{and}
  \bibinfo{author}{\bibfnamefont{B.~J.} \bibnamefont{LeRoy}},
  \bibinfo{journal}{Physical Review B} \textbf{\bibinfo{volume}{79}},
  \bibinfo{pages}{205411} (\bibinfo{year}{2009}).

\bibitem[{\citenamefont{Norris and Vemula}(1995)}]{norris1995scattering}
\bibinfo{author}{\bibfnamefont{A.}~\bibnamefont{Norris}} \bibnamefont{and}
  \bibinfo{author}{\bibfnamefont{C.}~\bibnamefont{Vemula}},
  \bibinfo{journal}{Journal of Sound and Vibration}
  \textbf{\bibinfo{volume}{181}}, \bibinfo{pages}{115} (\bibinfo{year}{1995}).

\bibitem[{\citenamefont{John et~al.}(1983)\citenamefont{John, Sompolinsky, and
  Stephen}}]{john1983localization}
\bibinfo{author}{\bibfnamefont{S.}~\bibnamefont{John}},
  \bibinfo{author}{\bibfnamefont{H.}~\bibnamefont{Sompolinsky}},
  \bibnamefont{and} \bibinfo{author}{\bibfnamefont{M.~J.}
  \bibnamefont{Stephen}}, \bibinfo{journal}{Physical Review B}
  \textbf{\bibinfo{volume}{27}}, \bibinfo{pages}{5592} (\bibinfo{year}{1983}).

\bibitem[{\citenamefont{Kirkpatrick}(1985)}]{kirkpatrick1985localization}
\bibinfo{author}{\bibfnamefont{T.}~\bibnamefont{Kirkpatrick}},
  \bibinfo{journal}{Physical Review B} \textbf{\bibinfo{volume}{31}},
  \bibinfo{pages}{5746} (\bibinfo{year}{1985}).

\bibitem[{\citenamefont{Akkermans and Maynard}(1985)}]{akkermans1985weak}
\bibinfo{author}{\bibfnamefont{E.}~\bibnamefont{Akkermans}} \bibnamefont{and}
  \bibinfo{author}{\bibfnamefont{R.}~\bibnamefont{Maynard}},
  \bibinfo{journal}{Physical Review B} \textbf{\bibinfo{volume}{32}},
  \bibinfo{pages}{7850} (\bibinfo{year}{1985}).

\bibitem[{\citenamefont{Sepehrinia et~al.}(2008)\citenamefont{Sepehrinia,
  Tabar, and Sahimi}}]{sepehrinia2008numerical}
\bibinfo{author}{\bibfnamefont{R.}~\bibnamefont{Sepehrinia}},
  \bibinfo{author}{\bibfnamefont{M.~R.~R.} \bibnamefont{Tabar}},
  \bibnamefont{and} \bibinfo{author}{\bibfnamefont{M.}~\bibnamefont{Sahimi}},
  \bibinfo{journal}{Physical Review B} \textbf{\bibinfo{volume}{78}},
  \bibinfo{pages}{024207} (\bibinfo{year}{2008}).

\bibitem[{\citenamefont{Monthus and Garel}(2010)}]{monthus2010anderson}
\bibinfo{author}{\bibfnamefont{C.}~\bibnamefont{Monthus}} \bibnamefont{and}
  \bibinfo{author}{\bibfnamefont{T.}~\bibnamefont{Garel}},
  \bibinfo{journal}{Physical Review B} \textbf{\bibinfo{volume}{81}},
  \bibinfo{pages}{224208} (\bibinfo{year}{2010}).

\bibitem[{\citenamefont{Conley et~al.}(2011)\citenamefont{Conley, Lavrik,
  Prasai, and Bolotin}}]{conley2011graphene}
\bibinfo{author}{\bibfnamefont{H.}~\bibnamefont{Conley}},
  \bibinfo{author}{\bibfnamefont{N.~V.} \bibnamefont{Lavrik}},
  \bibinfo{author}{\bibfnamefont{D.}~\bibnamefont{Prasai}}, \bibnamefont{and}
  \bibinfo{author}{\bibfnamefont{K.~I.} \bibnamefont{Bolotin}},
  \bibinfo{journal}{Nano Letters} \textbf{\bibinfo{volume}{11}},
  \bibinfo{pages}{4748} (\bibinfo{year}{2011}).

\bibitem[{\citenamefont{Sabio et~al.}(2008)\citenamefont{Sabio, Seoanez,
  Fratini, Guinea, Neto, and Sols}}]{sabio2008electrostatic}
\bibinfo{author}{\bibfnamefont{J.}~\bibnamefont{Sabio}},
  \bibinfo{author}{\bibfnamefont{C.}~\bibnamefont{Seoanez}},
  \bibinfo{author}{\bibfnamefont{S.}~\bibnamefont{Fratini}},
  \bibinfo{author}{\bibfnamefont{F.}~\bibnamefont{Guinea}},
  \bibinfo{author}{\bibfnamefont{A.~C.} \bibnamefont{Neto}}, \bibnamefont{and}
  \bibinfo{author}{\bibfnamefont{F.}~\bibnamefont{Sols}},
  \bibinfo{journal}{Physical Review B} \textbf{\bibinfo{volume}{77}},
  \bibinfo{pages}{195409} (\bibinfo{year}{2008}).

\bibitem[{\citenamefont{Kusminskiy et~al.}(2011)\citenamefont{Kusminskiy,
  Campbell, Neto, and Guinea}}]{kusminskiy2011pinning}
\bibinfo{author}{\bibfnamefont{S.~V.} \bibnamefont{Kusminskiy}},
  \bibinfo{author}{\bibfnamefont{D.}~\bibnamefont{Campbell}},
  \bibinfo{author}{\bibfnamefont{A.~C.} \bibnamefont{Neto}}, \bibnamefont{and}
  \bibinfo{author}{\bibfnamefont{F.}~\bibnamefont{Guinea}},
  \bibinfo{journal}{Physical Review B} \textbf{\bibinfo{volume}{83}},
  \bibinfo{pages}{165405} (\bibinfo{year}{2011}).

\bibitem[{\citenamefont{Meyer et~al.}(2007)\citenamefont{Meyer, Geim,
  Katsnelson, Novoselov, Booth, and Roth}}]{meyer2007suspended}
\bibinfo{author}{\bibfnamefont{J.~C.} \bibnamefont{Meyer}},
  \bibinfo{author}{\bibfnamefont{A.~K.} \bibnamefont{Geim}},
  \bibinfo{author}{\bibfnamefont{M.~I.} \bibnamefont{Katsnelson}},
  \bibinfo{author}{\bibfnamefont{K.~S.} \bibnamefont{Novoselov}},
  \bibinfo{author}{\bibfnamefont{T.~J.} \bibnamefont{Booth}}, \bibnamefont{and}
  \bibinfo{author}{\bibfnamefont{S.}~\bibnamefont{Roth}},
  \bibinfo{journal}{Nature} \textbf{\bibinfo{volume}{446}}, \bibinfo{pages}{60}
  (\bibinfo{year}{2007}).

\bibitem[{\citenamefont{Laitinen et~al.}(2014)\citenamefont{Laitinen, Oksanen,
  Fay, Cox, Tomi, Virtanen, and Hakonen}}]{laitinen2014}
\bibinfo{author}{\bibfnamefont{A.}~\bibnamefont{Laitinen}},
  \bibinfo{author}{\bibfnamefont{M.}~\bibnamefont{Oksanen}},
  \bibinfo{author}{\bibfnamefont{A.}~\bibnamefont{Fay}},
  \bibinfo{author}{\bibfnamefont{D.}~\bibnamefont{Cox}},
  \bibinfo{author}{\bibfnamefont{M.}~\bibnamefont{Tomi}},
  \bibinfo{author}{\bibfnamefont{P.}~\bibnamefont{Virtanen}}, \bibnamefont{and}
  \bibinfo{author}{\bibfnamefont{P.~J.} \bibnamefont{Hakonen}},
  \bibinfo{journal}{Nano Letters} \textbf{\bibinfo{volume}{14}},
  \bibinfo{pages}{3009} (\bibinfo{year}{2014}).

\bibitem[{\citenamefont{Morozov et~al.}(2008)\citenamefont{Morozov, Novoselov,
  Katsnelson, Schedin, Elias, Jaszczak, and Geim}}]{morozov2008giant}
\bibinfo{author}{\bibfnamefont{S.}~\bibnamefont{Morozov}},
  \bibinfo{author}{\bibfnamefont{K.}~\bibnamefont{Novoselov}},
  \bibinfo{author}{\bibfnamefont{M.}~\bibnamefont{Katsnelson}},
  \bibinfo{author}{\bibfnamefont{F.}~\bibnamefont{Schedin}},
  \bibinfo{author}{\bibfnamefont{D.}~\bibnamefont{Elias}},
  \bibinfo{author}{\bibfnamefont{J.~A.} \bibnamefont{Jaszczak}},
  \bibnamefont{and} \bibinfo{author}{\bibfnamefont{A.}~\bibnamefont{Geim}},
  \bibinfo{journal}{Physical Review Letters} \textbf{\bibinfo{volume}{100}},
  \bibinfo{pages}{016602} (\bibinfo{year}{2008}).

\bibitem[{\citenamefont{Seol et~al.}(2010)\citenamefont{Seol, Jo, Moore,
  Lindsay, Aitken, Pettes, Li, Yao, Huang, Broido et~al.}}]{seol2010two}
\bibinfo{author}{\bibfnamefont{J.~H.} \bibnamefont{Seol}},
  \bibinfo{author}{\bibfnamefont{I.}~\bibnamefont{Jo}},
  \bibinfo{author}{\bibfnamefont{A.~L.} \bibnamefont{Moore}},
  \bibinfo{author}{\bibfnamefont{L.}~\bibnamefont{Lindsay}},
  \bibinfo{author}{\bibfnamefont{Z.~H.} \bibnamefont{Aitken}},
  \bibinfo{author}{\bibfnamefont{M.~T.} \bibnamefont{Pettes}},
  \bibinfo{author}{\bibfnamefont{X.}~\bibnamefont{Li}},
  \bibinfo{author}{\bibfnamefont{Z.}~\bibnamefont{Yao}},
  \bibinfo{author}{\bibfnamefont{R.}~\bibnamefont{Huang}},
  \bibinfo{author}{\bibfnamefont{D.}~\bibnamefont{Broido}},
  \bibnamefont{et~al.}, \bibinfo{journal}{Science}
  \textbf{\bibinfo{volume}{328}}, \bibinfo{pages}{213} (\bibinfo{year}{2010}).

\bibitem[{\citenamefont{Pumarol et~al.}(2012)\citenamefont{Pumarol, Rosamond,
  Tovee, Petty, Zeze, Falko, and Kolosov}}]{pumarol2012imaging}
\bibinfo{author}{\bibfnamefont{M.~E.} \bibnamefont{Pumarol}},
  \bibinfo{author}{\bibfnamefont{M.~C.} \bibnamefont{Rosamond}},
  \bibinfo{author}{\bibfnamefont{P.}~\bibnamefont{Tovee}},
  \bibinfo{author}{\bibfnamefont{M.~C.} \bibnamefont{Petty}},
  \bibinfo{author}{\bibfnamefont{D.~A.} \bibnamefont{Zeze}},
  \bibinfo{author}{\bibfnamefont{V.}~\bibnamefont{Falko}}, \bibnamefont{and}
  \bibinfo{author}{\bibfnamefont{O.~V.} \bibnamefont{Kolosov}},
  \bibinfo{journal}{Nano Letters} \textbf{\bibinfo{volume}{12}},
  \bibinfo{pages}{2906} (\bibinfo{year}{2012}).

\bibitem[{\citenamefont{He et~al.}(2013)\citenamefont{He, Chen, Yu, Ouyang, and
  Yang}}]{he2013anomalous}
\bibinfo{author}{\bibfnamefont{Y.}~\bibnamefont{He}},
  \bibinfo{author}{\bibfnamefont{W.}~\bibnamefont{Chen}},
  \bibinfo{author}{\bibfnamefont{W.}~\bibnamefont{Yu}},
  \bibinfo{author}{\bibfnamefont{G.}~\bibnamefont{Ouyang}}, \bibnamefont{and}
  \bibinfo{author}{\bibfnamefont{G.}~\bibnamefont{Yang}},
  \bibinfo{journal}{Scientific Reports} \textbf{\bibinfo{volume}{3}}
  (\bibinfo{year}{2013}).

\bibitem[{\citenamefont{Abrahams et~al.}(1979)\citenamefont{Abrahams, Anderson,
  Licciardello, and Ramakrishnan}}]{abrahams1979scaling}
\bibinfo{author}{\bibfnamefont{E.}~\bibnamefont{Abrahams}},
  \bibinfo{author}{\bibfnamefont{P.}~\bibnamefont{Anderson}},
  \bibinfo{author}{\bibfnamefont{D.}~\bibnamefont{Licciardello}},
  \bibnamefont{and}
  \bibinfo{author}{\bibfnamefont{T.}~\bibnamefont{Ramakrishnan}},
  \bibinfo{journal}{Physical Review Letters} \textbf{\bibinfo{volume}{42}},
  \bibinfo{pages}{673} (\bibinfo{year}{1979}).

\bibitem[{\citenamefont{Mirlin}(2000)}]{mirlin2000statistics}
\bibinfo{author}{\bibfnamefont{A.~D.} \bibnamefont{Mirlin}},
  \bibinfo{journal}{Physics Reports} \textbf{\bibinfo{volume}{326}},
  \bibinfo{pages}{259} (\bibinfo{year}{2000}).

\bibitem[{\citenamefont{Evers and Mirlin}(2008)}]{evers2008anderson}
\bibinfo{author}{\bibfnamefont{F.}~\bibnamefont{Evers}} \bibnamefont{and}
  \bibinfo{author}{\bibfnamefont{A.~D.} \bibnamefont{Mirlin}},
  \bibinfo{journal}{Reviews of Modern Physics} \textbf{\bibinfo{volume}{80}},
  \bibinfo{pages}{1355} (\bibinfo{year}{2008}).

\bibitem[{\citenamefont{Wegner}(1980)}]{Wegner1980}
\bibinfo{author}{\bibfnamefont{F.}~\bibnamefont{Wegner}},
  \bibinfo{journal}{Zeitschrift Fur Physik B-Condensed Matter}
  \textbf{\bibinfo{volume}{36}}, \bibinfo{pages}{209} (\bibinfo{year}{1980}).

\bibitem[{\citenamefont{Fal’ko and Efetov}(1995)}]{Fal'koEfetovPRB52}
\bibinfo{author}{\bibfnamefont{V.~I.} \bibnamefont{Fal’ko}} \bibnamefont{and}
  \bibinfo{author}{\bibfnamefont{K.}~\bibnamefont{Efetov}},
  \bibinfo{journal}{Physical Review B} \textbf{\bibinfo{volume}{52}},
  \bibinfo{pages}{17413} (\bibinfo{year}{1995}).

\bibitem[{\citenamefont{Kravtsov and Lerner}(1995)}]{kravtsov1995level}
\bibinfo{author}{\bibfnamefont{V.~E.} \bibnamefont{Kravtsov}} \bibnamefont{and}
  \bibinfo{author}{\bibfnamefont{I.~V.} \bibnamefont{Lerner}},
  \bibinfo{journal}{Physical Review Letters} \textbf{\bibinfo{volume}{74}},
  \bibinfo{pages}{2563} (\bibinfo{year}{1995}).

\bibitem[{\citenamefont{Braun and Montambaux}(1995)}]{braun1995spectral}
\bibinfo{author}{\bibfnamefont{D.}~\bibnamefont{Braun}} \bibnamefont{and}
  \bibinfo{author}{\bibfnamefont{G.}~\bibnamefont{Montambaux}},
  \bibinfo{journal}{Physical Review B} \textbf{\bibinfo{volume}{52}},
  \bibinfo{pages}{13903} (\bibinfo{year}{1995}).

\bibitem[{\citenamefont{Fyodorov and Mirlin}(1995)}]{fyodorov1995mesoscopic}
\bibinfo{author}{\bibfnamefont{Y.~V.} \bibnamefont{Fyodorov}} \bibnamefont{and}
  \bibinfo{author}{\bibfnamefont{A.~D.} \bibnamefont{Mirlin}},
  \bibinfo{journal}{Physical Review B} \textbf{\bibinfo{volume}{51}},
  \bibinfo{pages}{13403} (\bibinfo{year}{1995}).

\bibitem[{\citenamefont{Nelson and Peliti}(1987)}]{NelsonPeliti1987}
\bibinfo{author}{\bibfnamefont{D.}~\bibnamefont{Nelson}} \bibnamefont{and}
  \bibinfo{author}{\bibfnamefont{L.}~\bibnamefont{Peliti}},
  \bibinfo{journal}{J. Phys.(Paris)} \textbf{\bibinfo{volume}{48}},
  \bibinfo{pages}{1085} (\bibinfo{year}{1987}).

\bibitem[{\citenamefont{David and Guitter}(1988)}]{DavidGuitter1988}
\bibinfo{author}{\bibfnamefont{F.}~\bibnamefont{David}} \bibnamefont{and}
  \bibinfo{author}{\bibfnamefont{E.}~\bibnamefont{Guitter}},
  \bibinfo{journal}{EPL (Europhysics Letters)} \textbf{\bibinfo{volume}{5}},
  \bibinfo{pages}{709} (\bibinfo{year}{1988}).

\bibitem[{\citenamefont{Le~Doussal and Radzihovsky}(1992)}]{Doussal1992}
\bibinfo{author}{\bibfnamefont{P.}~\bibnamefont{Le~Doussal}} \bibnamefont{and}
  \bibinfo{author}{\bibfnamefont{L.}~\bibnamefont{Radzihovsky}},
  \bibinfo{journal}{Physical Review Letters} \textbf{\bibinfo{volume}{69}},
  \bibinfo{pages}{1209} (\bibinfo{year}{1992}).

\bibitem[{\citenamefont{Nelson et~al.}(2004)\citenamefont{Nelson, Piran, and
  Weinberg}}]{Nelsonbook}
\bibinfo{author}{\bibfnamefont{D.}~\bibnamefont{Nelson}},
  \bibinfo{author}{\bibfnamefont{T.}~\bibnamefont{Piran}}, \bibnamefont{and}
  \bibinfo{author}{\bibfnamefont{S.}~\bibnamefont{Weinberg}},
  \emph{\bibinfo{title}{Statistical mechanics of membranes and surfaces}}
  (\bibinfo{publisher}{World Scientific}, \bibinfo{year}{2004}).

\bibitem[{\citenamefont{Le~Doussal and
  Radzihovsky}(2017)}]{LeDoussalRadzihovsky2017}
\bibinfo{author}{\bibfnamefont{P.}~\bibnamefont{Le~Doussal}} \bibnamefont{and}
  \bibinfo{author}{\bibfnamefont{L.}~\bibnamefont{Radzihovsky}},
  \bibinfo{journal}{Annals of Physics}  (\bibinfo{year}{2017}).

\bibitem[{\citenamefont{Song et~al.}(2012)\citenamefont{Song, Reizer, and
  Levitov}}]{song2012}
\bibinfo{author}{\bibfnamefont{J.~C.} \bibnamefont{Song}},
  \bibinfo{author}{\bibfnamefont{M.~Y.} \bibnamefont{Reizer}},
  \bibnamefont{and} \bibinfo{author}{\bibfnamefont{L.~S.}
  \bibnamefont{Levitov}}, \bibinfo{journal}{Physical Review Letters}
  \textbf{\bibinfo{volume}{109}}, \bibinfo{pages}{106602}
  (\bibinfo{year}{2012}).

\bibitem[{\citenamefont{Los et~al.}(2009)\citenamefont{Los, Katsnelson, Yazyev,
  Zakharchenko, and Fasolino}}]{los2009}
\bibinfo{author}{\bibfnamefont{J.}~\bibnamefont{Los}},
  \bibinfo{author}{\bibfnamefont{M.~I.} \bibnamefont{Katsnelson}},
  \bibinfo{author}{\bibfnamefont{O.}~\bibnamefont{Yazyev}},
  \bibinfo{author}{\bibfnamefont{K.}~\bibnamefont{Zakharchenko}},
  \bibnamefont{and} \bibinfo{author}{\bibfnamefont{A.}~\bibnamefont{Fasolino}},
  \bibinfo{journal}{Physical Review B} \textbf{\bibinfo{volume}{80}},
  \bibinfo{pages}{121405} (\bibinfo{year}{2009}).

\bibitem[{\citenamefont{Bowick et~al.}(2017)\citenamefont{Bowick,
  Ko{\v{s}}mrlj, Nelson, and Sknepnek}}]{bowick2017non-Hookean}
\bibinfo{author}{\bibfnamefont{M.~J.} \bibnamefont{Bowick}},
  \bibinfo{author}{\bibfnamefont{A.}~\bibnamefont{Ko{\v{s}}mrlj}},
  \bibinfo{author}{\bibfnamefont{D.~R.} \bibnamefont{Nelson}},
  \bibnamefont{and} \bibinfo{author}{\bibfnamefont{R.}~\bibnamefont{Sknepnek}},
  \bibinfo{journal}{Physical Review B} \textbf{\bibinfo{volume}{95}},
  \bibinfo{pages}{104109} (\bibinfo{year}{2017}).

\bibitem[{\citenamefont{Blees et~al.}(2015)\citenamefont{Blees, Barnard, Rose,
  Roberts, McGill, Huang, Ruyack, Kevek, Kobrin, Muller et~al.}}]{blees2015}
\bibinfo{author}{\bibfnamefont{M.~K.} \bibnamefont{Blees}},
  \bibinfo{author}{\bibfnamefont{A.~W.} \bibnamefont{Barnard}},
  \bibinfo{author}{\bibfnamefont{P.~A.} \bibnamefont{Rose}},
  \bibinfo{author}{\bibfnamefont{S.~P.} \bibnamefont{Roberts}},
  \bibinfo{author}{\bibfnamefont{K.~L.} \bibnamefont{McGill}},
  \bibinfo{author}{\bibfnamefont{P.~Y.} \bibnamefont{Huang}},
  \bibinfo{author}{\bibfnamefont{A.~R.} \bibnamefont{Ruyack}},
  \bibinfo{author}{\bibfnamefont{J.~W.} \bibnamefont{Kevek}},
  \bibinfo{author}{\bibfnamefont{B.}~\bibnamefont{Kobrin}},
  \bibinfo{author}{\bibfnamefont{D.~A.} \bibnamefont{Muller}},
  \bibnamefont{et~al.}, \bibinfo{journal}{Nature}
  \textbf{\bibinfo{volume}{524}}, \bibinfo{pages}{204} (\bibinfo{year}{2015}).

\bibitem[{\citenamefont{Nicholl et~al.}(2015)\citenamefont{Nicholl, Conley,
  Lavrik, Vlassiouk, Puzyrev, Sreenivas, Pantelides, and
  Bolotin}}]{nicholl2015}
\bibinfo{author}{\bibfnamefont{R.~J.} \bibnamefont{Nicholl}},
  \bibinfo{author}{\bibfnamefont{H.~J.} \bibnamefont{Conley}},
  \bibinfo{author}{\bibfnamefont{N.~V.} \bibnamefont{Lavrik}},
  \bibinfo{author}{\bibfnamefont{I.}~\bibnamefont{Vlassiouk}},
  \bibinfo{author}{\bibfnamefont{Y.~S.} \bibnamefont{Puzyrev}},
  \bibinfo{author}{\bibfnamefont{V.~P.} \bibnamefont{Sreenivas}},
  \bibinfo{author}{\bibfnamefont{S.~T.} \bibnamefont{Pantelides}},
  \bibnamefont{and} \bibinfo{author}{\bibfnamefont{K.~I.}
  \bibnamefont{Bolotin}}, \bibinfo{journal}{Nature Communications}
  \textbf{\bibinfo{volume}{6}}, \bibinfo{pages}{8789} (\bibinfo{year}{2015}).

\bibitem[{\citenamefont{Ko{\v{s}}mrlj and Nelson}(2016)}]{kosmrlj2016}
\bibinfo{author}{\bibfnamefont{A.}~\bibnamefont{Ko{\v{s}}mrlj}}
  \bibnamefont{and} \bibinfo{author}{\bibfnamefont{D.~R.}
  \bibnamefont{Nelson}}, \bibinfo{journal}{Physical Review B}
  \textbf{\bibinfo{volume}{93}}, \bibinfo{pages}{125431}
  (\bibinfo{year}{2016}).

\bibitem[{\citenamefont{Los et~al.}(2016)\citenamefont{Los, Fasolino, and
  Katsnelson}}]{los2016}
\bibinfo{author}{\bibfnamefont{J.}~\bibnamefont{Los}},
  \bibinfo{author}{\bibfnamefont{A.}~\bibnamefont{Fasolino}}, \bibnamefont{and}
  \bibinfo{author}{\bibfnamefont{M.}~\bibnamefont{Katsnelson}},
  \bibinfo{journal}{Physical Review Letters} \textbf{\bibinfo{volume}{116}},
  \bibinfo{pages}{015901} (\bibinfo{year}{2016}).

\bibitem[{\citenamefont{Kats and Lebedev}(2014)}]{KatsLebedev2014}
\bibinfo{author}{\bibfnamefont{E.}~\bibnamefont{Kats}} \bibnamefont{and}
  \bibinfo{author}{\bibfnamefont{V.}~\bibnamefont{Lebedev}},
  \bibinfo{journal}{Physical Review B} \textbf{\bibinfo{volume}{89}},
  \bibinfo{pages}{125433} (\bibinfo{year}{2014}).

\bibitem[{\citenamefont{Borzenets et~al.}(2013)\citenamefont{Borzenets, Coskun,
  Mebrahtu, Bomze, Smirnov, and Finkelstein}}]{borzenets2013phononbottleneck}
\bibinfo{author}{\bibfnamefont{I.}~\bibnamefont{Borzenets}},
  \bibinfo{author}{\bibfnamefont{U.}~\bibnamefont{Coskun}},
  \bibinfo{author}{\bibfnamefont{H.}~\bibnamefont{Mebrahtu}},
  \bibinfo{author}{\bibfnamefont{Y.~V.} \bibnamefont{Bomze}},
  \bibinfo{author}{\bibfnamefont{A.}~\bibnamefont{Smirnov}}, \bibnamefont{and}
  \bibinfo{author}{\bibfnamefont{G.}~\bibnamefont{Finkelstein}},
  \bibinfo{journal}{Physical Review Letters} \textbf{\bibinfo{volume}{111}},
  \bibinfo{pages}{027001} (\bibinfo{year}{2013}).

\bibitem[{\citenamefont{Fong et~al.}(2013)\citenamefont{Fong, Wollman, Ravi,
  Chen, Clerk, Shaw, Leduc, and Schwab}}]{KinChungFong2013fig3}
\bibinfo{author}{\bibfnamefont{K.~C.} \bibnamefont{Fong}},
  \bibinfo{author}{\bibfnamefont{E.~E.} \bibnamefont{Wollman}},
  \bibinfo{author}{\bibfnamefont{H.}~\bibnamefont{Ravi}},
  \bibinfo{author}{\bibfnamefont{W.}~\bibnamefont{Chen}},
  \bibinfo{author}{\bibfnamefont{A.~A.} \bibnamefont{Clerk}},
  \bibinfo{author}{\bibfnamefont{M.}~\bibnamefont{Shaw}},
  \bibinfo{author}{\bibfnamefont{H.}~\bibnamefont{Leduc}}, \bibnamefont{and}
  \bibinfo{author}{\bibfnamefont{K.}~\bibnamefont{Schwab}},
  \bibinfo{journal}{Physical Review X} \textbf{\bibinfo{volume}{3}},
  \bibinfo{pages}{041008} (\bibinfo{year}{2013}).

\bibitem[{\citenamefont{McKitterick et~al.}(2016)\citenamefont{McKitterick,
  Prober, and Rooks}}]{prober2016}
\bibinfo{author}{\bibfnamefont{C.~B.} \bibnamefont{McKitterick}},
  \bibinfo{author}{\bibfnamefont{D.~E.} \bibnamefont{Prober}},
  \bibnamefont{and} \bibinfo{author}{\bibfnamefont{M.~J.} \bibnamefont{Rooks}},
  \bibinfo{journal}{Physical Review B} \textbf{\bibinfo{volume}{93}},
  \bibinfo{pages}{075410} (\bibinfo{year}{2016}).

\bibitem[{\citenamefont{Bistritzer and MacDonald}(2009)}]{bistritzer2009}
\bibinfo{author}{\bibfnamefont{R.}~\bibnamefont{Bistritzer}} \bibnamefont{and}
  \bibinfo{author}{\bibfnamefont{A.}~\bibnamefont{MacDonald}},
  \bibinfo{journal}{Physical Review Letters} \textbf{\bibinfo{volume}{102}},
  \bibinfo{pages}{206410} (\bibinfo{year}{2009}).

\bibitem[{\citenamefont{Chen and Clerk}(2012)}]{chen2012}
\bibinfo{author}{\bibfnamefont{W.}~\bibnamefont{Chen}} \bibnamefont{and}
  \bibinfo{author}{\bibfnamefont{A.~A.} \bibnamefont{Clerk}},
  \bibinfo{journal}{Physical Review B} \textbf{\bibinfo{volume}{86}},
  \bibinfo{pages}{125443} (\bibinfo{year}{2012}).

\bibitem[{\citenamefont{Schwiete and Finkel'stein}(2014)}]{Schwiete2014}
\bibinfo{author}{\bibfnamefont{G.}~\bibnamefont{Schwiete}} \bibnamefont{and}
  \bibinfo{author}{\bibfnamefont{A.~M.} \bibnamefont{Finkel'stein}},
  \bibinfo{journal}{Physical Review B} \textbf{\bibinfo{volume}{90}},
  \bibinfo{pages}{060201} (\bibinfo{year}{2014}).

\bibitem[{\citenamefont{Schwiete and Finkel'stein}(2016)}]{Schwiete2016a}
\bibinfo{author}{\bibfnamefont{G.}~\bibnamefont{Schwiete}} \bibnamefont{and}
  \bibinfo{author}{\bibfnamefont{A.~M.} \bibnamefont{Finkel'stein}},
  \bibinfo{journal}{Physical Review B} \textbf{\bibinfo{volume}{93}},
  \bibinfo{pages}{115121} (\bibinfo{year}{2016}).

\bibitem[{\citenamefont{Schwiete and Finkel’stein}(2016)}]{Schwiete2016b}
\bibinfo{author}{\bibfnamefont{G.}~\bibnamefont{Schwiete}} \bibnamefont{and}
  \bibinfo{author}{\bibfnamefont{A.~M.} \bibnamefont{Finkel’stein}},
  \bibinfo{journal}{Journal of Experimental and Theoretical Physics}
  \textbf{\bibinfo{volume}{122}}, \bibinfo{pages}{567} (\bibinfo{year}{2016}).

\bibitem[{\citenamefont{Radzihovsky and Nelson}(1991)}]{Radzihovsky1991}
\bibinfo{author}{\bibfnamefont{L.}~\bibnamefont{Radzihovsky}} \bibnamefont{and}
  \bibinfo{author}{\bibfnamefont{D.~R.} \bibnamefont{Nelson}},
  \bibinfo{journal}{Physical Review A} \textbf{\bibinfo{volume}{44}},
  \bibinfo{pages}{3525} (\bibinfo{year}{1991}).

\bibitem[{\citenamefont{Morse and Lubensky}(1992)}]{Morse1992}
\bibinfo{author}{\bibfnamefont{D.~C.} \bibnamefont{Morse}} \bibnamefont{and}
  \bibinfo{author}{\bibfnamefont{T.~C.} \bibnamefont{Lubensky}},
  \bibinfo{journal}{Physical Review A} \textbf{\bibinfo{volume}{46}},
  \bibinfo{pages}{1751} (\bibinfo{year}{1992}).

\bibitem[{\citenamefont{Le~Doussal and Radzihovsky}(1993)}]{leDoussal1993}
\bibinfo{author}{\bibfnamefont{P.}~\bibnamefont{Le~Doussal}} \bibnamefont{and}
  \bibinfo{author}{\bibfnamefont{L.}~\bibnamefont{Radzihovsky}},
  \bibinfo{journal}{Phys. Rev. B} \textbf{\bibinfo{volume}{48}},
  \bibinfo{pages}{3548} (\bibinfo{year}{1993}).

\bibitem[{\citenamefont{Gornyi et~al.}(2012)\citenamefont{Gornyi, Kachorovskii,
  and Mirlin}}]{gornyi12}
\bibinfo{author}{\bibfnamefont{I.}~\bibnamefont{Gornyi}},
  \bibinfo{author}{\bibfnamefont{V.~Y.} \bibnamefont{Kachorovskii}},
  \bibnamefont{and} \bibinfo{author}{\bibfnamefont{A.}~\bibnamefont{Mirlin}},
  \bibinfo{journal}{Physical Review B} \textbf{\bibinfo{volume}{86}},
  \bibinfo{pages}{165413} (\bibinfo{year}{2012}).

\bibitem[{\citenamefont{Ko{\v{s}}mrlj and Nelson}(2013)}]{kosmrlj2013}
\bibinfo{author}{\bibfnamefont{A.}~\bibnamefont{Ko{\v{s}}mrlj}}
  \bibnamefont{and} \bibinfo{author}{\bibfnamefont{D.~R.}
  \bibnamefont{Nelson}}, \bibinfo{journal}{Physical Review E}
  \textbf{\bibinfo{volume}{88}}, \bibinfo{pages}{012136}
  (\bibinfo{year}{2013}).

\bibitem[{\citenamefont{Gornyi et~al.}(2015)\citenamefont{Gornyi, Kachorovskii,
  and Mirlin}}]{gornyi2015rippling}
\bibinfo{author}{\bibfnamefont{I.}~\bibnamefont{Gornyi}},
  \bibinfo{author}{\bibfnamefont{V.~Y.} \bibnamefont{Kachorovskii}},
  \bibnamefont{and} \bibinfo{author}{\bibfnamefont{A.}~\bibnamefont{Mirlin}},
  \bibinfo{journal}{Physical Review B} \textbf{\bibinfo{volume}{92}},
  \bibinfo{pages}{155428} (\bibinfo{year}{2015}).

\bibitem[{\citenamefont{L{\'o}pez-Pol{\'\i}n
  et~al.}(2015)\citenamefont{L{\'o}pez-Pol{\'\i}n, G{\'o}mez-Navarro, Parente,
  Guinea, Katsnelson, P{\'e}rez-Murano, and
  G{\'o}mez-Herrero}}]{Lopez-Polin2015}
\bibinfo{author}{\bibfnamefont{G.}~\bibnamefont{L{\'o}pez-Pol{\'\i}n}},
  \bibinfo{author}{\bibfnamefont{C.}~\bibnamefont{G{\'o}mez-Navarro}},
  \bibinfo{author}{\bibfnamefont{V.}~\bibnamefont{Parente}},
  \bibinfo{author}{\bibfnamefont{F.}~\bibnamefont{Guinea}},
  \bibinfo{author}{\bibfnamefont{M.~I.} \bibnamefont{Katsnelson}},
  \bibinfo{author}{\bibfnamefont{F.}~\bibnamefont{P{\'e}rez-Murano}},
  \bibnamefont{and}
  \bibinfo{author}{\bibfnamefont{J.}~\bibnamefont{G{\'o}mez-Herrero}},
  \bibinfo{journal}{Nature Physics} \textbf{\bibinfo{volume}{11}},
  \bibinfo{pages}{26} (\bibinfo{year}{2015}).

\bibitem[{\citenamefont{Tikhonov et~al.}(2014)\citenamefont{Tikhonov, Zhao, and
  Finkel’stein}}]{tikhonov2014dephasing}
\bibinfo{author}{\bibfnamefont{K.~S.} \bibnamefont{Tikhonov}},
  \bibinfo{author}{\bibfnamefont{W.~L.} \bibnamefont{Zhao}}, \bibnamefont{and}
  \bibinfo{author}{\bibfnamefont{A.~M.} \bibnamefont{Finkel’stein}},
  \bibinfo{journal}{Physical Review Letters} \textbf{\bibinfo{volume}{113}},
  \bibinfo{pages}{076601} (\bibinfo{year}{2014}).

\end{thebibliography}
\bigskip
\noindent \textbf{Acknowledgements:} The authors thank  Eva Andrei, Kirill Bolotin, Igor Gornyi, Alexander Mirlin, Daniel Prober, Valentin Kachorovskii and Eli Zeldov for useful discussions.

\noindent \textbf{Funding:} The work at the Landau Institute for Theoretical Physics (KT) was supported  by  the  Russian  Science  Foundation  under  the grant No.  14-42-00044.

\noindent A part 
of this research was conducted with the advanced computing resources provided by Texas A\&M High Performance Research Computing. The work at the Texas A$\&$M University (A.F. and W.Z.) is supported by the U.S. Department of Energy, Office of Basic Energy Sciences, Division of Materials Sciences and Engineering under Award DE-SC0014154.

\noindent \textbf{Author contributions:} W.Z. made computer simulations. All authors contributed substantially in the analysis of the data.
W.Z. and K.T. prepared figures. K.T and A.F. wrote the final version of the text.

\noindent \textbf{Competing interests:} The authors declare that they have no competing interests.

\noindent \textbf{Data availability:} All data needed to evaluate the conclusions in the paper are present in the
paper. Additional data related to this paper may be requested from the authors. Correspondence and requests for materials should be addressed to K.T. (email: tikhonov@itp.ac.ru)


\end{document}